# ORION'S VEIL – MAGNETIC FIELD STRENGTHS AND OTHER PROPERTIES OF A PDR IN FRONT OF THE TRAPEZIUM CLUSTER

T. H. Troland[1], W. M. Goss[2], C. L. Brogan[3], R. M. Crutcher[4], AND D. A. Roberts[5]

[1] Physics & Astronomy Department, University of Kentucky, Lexington, KY 40506, USA; troland@pa.uky.edu
[2] National Radio Astronomy Observatory, P.O. Box 0, Socorro, NM 87801, USA
[3] National Radio Astronomy Observatory, 520 Edgemont Rd, Charlottesville, VA 22903, USA
[4] Department of Astronomy, University of Illinois, Urbana-Champaign, IL 61801, USA
[5] Department of Physics & Astronomy and CIERA, Northwestern University, Evanston, IL 60208 USA

## ABSTRACT

We present an analysis of physical conditions in the Orion Veil, an atomic PDR that lies just in front ($\approx$ 2 pc) of the Trapezium stars of Orion. This region offers an unusual opportunity to study the properties of PDRs, including the magnetic field. We have obtained 21 cm H I and 18 cm (1665 and 1667 MHz) OH Zeeman effect data that yield images of the line-of-sight magnetic field strength $B_{los}$ in atomic and molecular regions of the Veil. We find $B_{los} \approx$ -50 to -75 µG in the atomic gas across much of the Veil (25" resolution); $B_{los} \approx$ -350 µG at one position in the molecular gas (40" resolution). The Veil has two principal H I velocity components. Magnetic and kinematical data suggest a close connection between these components. They may represent gas on either side of a shock wave preceding a weak-D ionization front. Magnetic fields in the Veil HI components are 3-5 times stronger than they are elsewhere in the ISM where $N$(H) and $n$(H) are comparable. The HI components are magnetically subcritical (magnetically dominated), like the CNM, although they are about 1 dex denser. Comparatively strong fields in the Veil HI components may have resulted from low turbulence conditions in the diffuse gas that gave rise to OMC-1. Strong fields may also be related to magnetostatic equilibrium that has developed in the Veil since star formation. We also consider the location of the Orion-S molecular core, proposing a location behind the main Orion H$^+$ region.

*Key words*: H II regions – ISM: individual objects (Orion Nebula, NGC 1976, M42, Orion A, Orion-S)
*Online-only material*: color figures
*Short title*: Orion Veil – Magnetic Fields

## 1. INTRODUCTION

Early-type stars irradiate their environments, creating photon-dominated regions (PDRs) with layers of ionized, atomic and molecular gas. This interaction between stars and the surrounding interstellar medium, in turn, profoundly affects the evolution of the region, triggering further star formation or else quenching that process by dissipating the clouds. The nearest locale in which these interactions can be studied is the Orion Nebula neighborhood (436 ± 20 pc, O'Dell & Henney 2008). In this region, there are three PDRs viewed from distinctly different perspectives. One PDR is the Orion Nebula itself (NGC 1976, M42) and the molecular cloud OMC-1 that lies behind it. This region, viewed face-on and lying behind the Trapezium Cluster, can be studied in optical, infrared and radio emission lines (e.g. Baldwin et al. 1991). The second Orion PDR is the Bright Bar. This region of bright optical emission along the southeast edge of the Orion Nebula is a PDR viewed nearly edge-on (*e.g.* Tielens et al. 1993; Hollenbach & Tielens 1997). Finally, the Orion Veil is a PDR viewed face on, like the Orion Nebula;

however, it lies in *front* of the Trapezium Cluster. Therefore, the Veil can be studied in optical and UV absorption lines against the Trapezium stars and in radio absorption lines against the nebular free-free emission. Abel et al. (2004) provide a simplified diagram of the Orion region, viewed sideways. O'Dell et al. (2009) provide a more detailed diagram, as do van der Werf et al. (2013, hereafter, vdWGO13). O'Dell (2001) and Ferland (2001) have reviewed observational data regarding the Orion Nebula and its environment.

Although the Veil contains a very small fraction of the mass of the Orion region, it offers an important laboratory for study of interstellar material associated with star formation. This importance comes from a rich body of observational data. These data include UV absorption lines from $H_2$ and from various atomic species (see Abel et al. 2004, 2006, 2016), optical Ca II and Na I absorption lines (Price et al. 2001; O'Dell et al. 1993), and radio absorption lines of 21 cm H I (e.g. vdWGO13), and 18 cm OH (this work). Abel et al. (2004, 2006, 2016) used observations of radio and UV absorption lines and spectral synthesis modeling to derive physical properties of the Veil such as temperature, density, and distance from the Trapezium stars.

In this study, we focus on the magnetic field of the Veil. The H I and OH lines are sensitive to the Zeeman effect in the atomic and molecular gas, respectively. Therefore, they offer an opportunity to image line-of-sight magnetic field strengths $B_{los}$ across the Veil via aperture synthesis techniques. Zeeman effect studies of the Veil go back to the early work of Verschuur (1969) and Brooks et al. (1971), both of whom detected the H I Zeeman effect with single dishes. Later, Troland et al. (1989) used the VLA to image the magnetic field in the Veil via the H I Zeeman effect. Troland et al. (1986) detected the 1667 MHz OH Zeeman effect with a single dish. The present VLA data provide higher H I sensitivity and spatial resolution than those of Troland et al. (1989), and they add Zeeman effect data for the 1665 and 1667 MHz OH lines.

To orient the reader, we cite several published images of the Orion region. The first is the spectacular HST-based optical image from Henney et al. (2007), reproduced here with modifications as Figure 1a. For clarity, we add star symbols to the figure to identify the four Trapezium stars. (See the figure caption for the meanings of other annotations.) Most of the features of the Orion region discussed in this study are apparent in Figure 1a. The names of the features are in general use; see O'Dell et al. 2009 and O'Dell & Harris 2010. The *Huygens region* is the brightest part of the Orion Nebula; it is about 6' (0.8 pc) across, and it appears in light colors centered upon the Trapezium stars. The *Bright Bar* (see above) is a 3' long linear feature marking the southeast edge of the Huygens region. *M43* is the small $H^+$ region located about 8' north northeast of the Trapezium stars at the very top of Figure 1a. M43 is excited by NU Ori, a B0-type star (Smith et al. 2005). The *Northeast Dark Lane* is the area of obvious extinction that separates M43 from the Huygens region. The *Dark Bay* is another region of obvious extinction that extends like a finger across the Huygens region from east to west, terminating about 1' northeast of the Trapezium stars. Finally, *Orion-S* is a dense molecular core discussed in several sections below. Orion-S is not visible in Figure 1a; however, the yellow circle centered about 1.5' southwest of the Trapezium stars denotes its location. Other notable Orion images are the 20 cm radio continuum and the HST Hα images of the Huygens region presented on the same scale by O'Dell & Yusef-Zadeh (2000). Comparison of these two images provides a sense of the optical extinction across the nebula, and O'Dell & Yusef-Zadeh use the images to construct an optical extinction image. They argue that optical extinction in this image comes principally from the Veil rather than from the main Orion $H^+$ region. In Figure 1b we present a finder chart for individual positions where we consider HI data (smaller circles) and OH data (larger circles). See Tables 2 & 3 as well as the discussion below.

## 2. OBSERVATIONS AND DATA ANALYSIS

Observations of the Zeeman effect in the 21 cm H I line and in the 18 cm OH main lines (1665 and 1667 MHz) were conducted with the Karl G. Jansky Very Large Array (VLA)[1]. H I data were obtained in the VLA C-array, while OH data were obtained in the D array. All lines were observed in absorption against the Orion Nebula radio continuum. Observational parameters are given in Tables 1a and 1b for the H I data and the OH data, respectively. The general techniques for HI and OH Zeeman observations and data analysis with the VLA have been described by previous authors (e.g. Roberts et al. 1993 and 1995; Brogan & Troland 2001; Sarma et al. 2000; Sarma et al. 2013). Nonetheless, for the sake of completeness, we review these techniques below. We note that all published VLA Zeeman observations, and those described here, were made prior to the VLA upgrade in 2012. As is conventional for galactic VLA observations, velocities cited in this work are in the kinematic LSR (LSRK) reference frame; heliocentric velocities toward Orion are 18.1 km s$^{-1}$ more positive than LSRK velocities. Henceforth, we refer to the LSRK reference frame as the LSR frame.

We begin with an explicit statement of polarization definitions used in this work. We use the IEEE standards for sense of circular polarization. That is, in right circular polarization (RCP) the electric vector rotates *clockwise* as a wave propagates *away from* the observer. We also use the IEEE definition of Stokes parameter V = RHC – LHC. With these definitions, a magnetic field directed *away from* the observer generates a Zeeman effect in which the line observed in LCH is higher in frequency than the line in RCP. Such fields are defined as *positive* in sign. All fields observed in the Orion Veil are *negative*, that is, directed toward the observer.

VLA Zeeman observations at 18 and 21 cm were made with dual native circular polarization receivers (RHC and LHC) on each telescope. Therefore, the construction of Stokes V profiles, necessary for detection of the Zeeman effect, requires subtraction of data from two independent receiver/IF systems with different bandpass shapes. The subtraction process introduces an instrumental bandpass shape into the Stokes V profiles. To mitigate this effect, we switched the RF transfer switches on each telescope every 10 minutes. This step inverts the senses of circular polarization sent to the two IF systems. When data from both switch positions are combined, bandpass differences downstream from the switches are eliminated. HI observations were made in the Two IF Mode of the original VLA correlator (one spectral line, two circular polarizations). OH observations were made in the Four IF Mode (two spectral lines, two circular polarizations). Finally, calibration sources for the HI observations were observed at frequencies displaced above and below the HI line by 1 MHz (210 km s$^{-1}$). Phase calibration solutions were then determined from the average of calibration scans at both displaced frequencies. This procedure ensured that galactic HI emission did not affect calibration source data.

Calibration and imaging of the HI and OH data were done with the NRAO AIPS package. Standard techniques were applied to calibrate RHC and LHC uv data from each transfer switch position (i.e. four data sets independently calibrated for each spectral line). Further AIPS analysis led to the production of Stokes I, Stokes V and optical depth image cubes for the HI line and for the two OH lines.  In addition, we took special steps to remove maser emission from the OH data, as described below.  These steps were taken to prevent images of the Stokes I and V OH thermal emission from being dominated (and dynamic range limited) by the strong maser emission.  Note that errors in absolute calibration of the data do not affect Zeeman measurements since magnetic field strengths are derived from ratios involving Stokes I and V.  However, the absolute VLA flux density scale is expected to accurate to about 5%.

---

[1] The National Radio Astronomy Observatory is a facility of the National Science Foundation operated under cooperative agreement by Associated Universities, Inc..

We used the AIPS task IMAGR to create Stokes I and Stokes V image cubes.  For the HI line, we created cleaned Stokes I image cubes from bandpass-corrected, continuum-subtracted uv data.  We chose a robustness of 1, intermediate between uniform and natural weighting of the uv data.  This choice was a compromise between the conflicting needs to achieve high angular resolution and low noise.  The HI Stokes V image cubes were created by IMAGR from uv data that had *not* been bandpass corrected.  With the transfer switching technique described above, the Stokes V bandpasses were acceptable without correction.  Moreover, bandpass correction adds noise to the Stokes V data, reducing sensitivity to the Zeeman effect.  Note that IMAGR calculates Stokes I/2 and Stokes V/2 images.

For each of the OH lines, we removed maser emission from the uv data cubes via a multi-step process.  We applied this process only to the maser channels, that is, channels with significant maser emission.  First, we created all-channel, bandpass-corrected, continuum-subtracted image cubes in RCP and LCP.  For each image cube, we performed self-calibration on the maser channels and reimaged each such channel.  Next we identified clean components in these channel images by specifying small clean boxes around each maser position.   Then we subtracted the contributions of the maser clean components from the uv data via UVSUB and reimaged the maser channels in RCP and LCP.   Finally, we replaced the original maser channels in the RCP and LCP image cubes with the newly maser-free maser channels.  Sarma et al. (2000) applied the same techniques to remove the effects of OH masers in NGC 6334.  To generate optical depth image cubes for HI and OH lines, we used task COMB to combine the continuum-subtracted and cleaned Stokes I cube with a cleaned continuum map constructed from line-free channels in the Stokes I profiles.  We were careful to clean the Stokes I cube and the continuum image in consistent ways to mitigate systematic errors in the optical depth calculations.

Once the Stokes I and V image cubes were produced for each line with AIPS tasks, we undertook the Zeeman analysis using the Multichannel Image Reconstruction Image Analysis and Display (MIRIAD) package with task ZEEMAP. This task performs a pixel-by-pixel fit of the Stokes V profile to the derivative of the Stokes I profile.  The fitting process, most recently described by Sarma et al. (2013), yields estimates of the line-of-sight magnetic field strength $B_{los}$ (and its sign), together with the error $\sigma(B_{los})$.  The fit for each pixel can be done over a specified range of spectral channels.  For the Orion HI data, we performed independent fits over two ranges of channels to estimate $B_{los}$ and $\sigma(B_{los})$ for each of two velocity components in the Veil (section 3.2).  As is customary in imaging VLA data, the pixel separations were chosen to be smaller than the synthesized beam widths.  (See Tables 1a and 1b.)  Therefore, values for $B_{los}$ and its error derived for adjacent pixels are not independent.  Rather, these values are averages over a synthesized beam centered on a particular pixel.

# 3. RESULTS

## 3.1. Radio Continuum

H I and OH absorption lines are observed against the radio free-free continuum of M42 and M43. An image of this continuum, derived by combining information from line-free channels in the Stokes I image cube, is a natural product of the data. Our 21 cm continuum image has a synthesized beamwidth of 25" (0.05 pc); our 18 cm (1665 and 1667 MHz) OH images have a synthesized beamwidth of 40" (0.09 pc). The peak brightness of our 21 cm image is 4.7 Jy beam$^{-1}$, equivalent to brightness temperature $T_B$ = 5100 K. We do not present our continuum images as separate figures. Equivalent or higher spatial resolution Orion continuum images at similar wavelengths are available in the literature. See, especially, the 21 cm continuum image of vdWGO13. See, also, the 20 cm image of O'Dell & Yusef-Zadeh (2000) and the wide field 330 MHz (90 cm) image of Subrahmanyan et al. (2001). We recover a total 21 cm continuum flux of 290 ± 15 Jy for M42. This value is to be compared with 335 ± 15 Jy found for M42 by vdWGO13. M43 contributes an additional 12 ± 2 Jy to our data, compared to 14 ± 2 Jy for vdWGO13. Errors reflect uncertainties in the allocation of flux between M42 and M43. See van der Werf & Goss (1989, hereafter, vdWG89) and vdWGO13 for additional information about the total Orion 21 cm continuum flux density.

## 3.2. H I Optical Depths

We constructed an H I optical depth image cube from the Stokes I image cube and the Stokes I continuum image. Toward the brightest part of M42, we measure peak H I optical depths as high as $\tau_{0,\,H\,I} \approx 5$. Over much of the Huygens region, we measure $\tau_{0,\,H\,I}$ of up to 4, and we measure H I optical depths out to the continuum brightness contour that is 3% of the peak. In general, H I optical depths increase from the southwest toward the northeast of the Veil, and the H I lines are saturated in the northeast sector of the image. Line saturation precludes measurement of $\tau_{H\,I} \propto N(H^0)/T_{ex}$, where $T_{ex}$ is the H I line excitation temperature. However, line saturation does not preclude detection of the H I Zeeman effect; see section 3.4.

The spatial variation of H I optical depths has already been described by other authors beginning with Lockhart & Goss (1978). In particular, vdWG89 observed H I absorption in the Veil with the VLA C-array. Although their velocity resolution was half that of the present observations, they were able to identify several velocity components. Principal among them are components A and B (their designations, adopted here). Components A and B have typical LSR center velocities of 5.0 and 1.3 km s$^{-1}$, respectively, and typical FWHM of 2.5 and 3.6 km s$^{-1}$, respectively (Table 3). Clearly, there are multiple layers of H$^0$ gas in the Veil. vdWG89 performed a Gaussian analysis of their H I optical depth data cube. From this analysis, they derived images of $N(H^0)/T_{ex}$ across the Veil for each component separately. They found that components A and B exist across the entire radio continuum source (M42 and M43), with peak optical depths generally higher in component A. However, an inspection of our more sensitive and higher velocity resolution H I optical depth profiles clearly reveals that the kinematical structure of the Veil is too complicated to be fully described by a two-component model. Nonetheless, the Gaussian decomposition done by vdWG89 provides a good qualitative idea of the distribution of H I velocity components A and B for which we have H I Zeeman effect images (section 3.4). vdWGO13 also present extensive VLA data on H I absorption and emission in the Veil. However, these authors concentrate on small scale structure, generally at velocities outside the range occupied by components A and B.

### 3.3. OH Optical Depths and Column Densities

The OH optical depth cubes show that OH absorption across the Veil is qualitatively different from H I absorption in two ways. For one, OH absorption is only detected in four isolated regions, not across the entire Veil. The second difference is that the OH profiles usually consist of a single velocity component, compared to at least two components in the HI profiles. OH absorption is *not* detected toward the Trapezium Cluster. This absence is expected, given the very low $H_2$ abundance toward the Trapezium stars ($N(H_2)/N(H^0) \approx 10^{-6.5}$) measured by UV absorption lines (Abel et al. 2016). The Veil along this line-of-sight is almost completely atomic owing to the very high stellar radiation field from the Trapezium.

Four regions of OH absorption appear in Figure 2. Here we present a color coded image of the velocity integrated 1667 MHz OH optical depth, in units of $[N(OH)/T_{ex}] \times 10^{14}$ cm$^{-2}$ K$^{-1}$, where $T_{ex}$ is the 1667 MHz excitation temperature. To calculate $N(OH)/T_{ex}$ from the velocity integrated optical depth profiles, we use the conversion factor given in Roberts et al. (1995), $C_{1667} = 2.28 \times 10^{14}$ cm$^{-2}$ K$^{-1}$ (km s$^{-1}$)$^{-1}$. This factor assumes LTE conditions. For reference, we overlay 1667 MHz continuum contours in the figure. These contours clearly show M42 as well as M43 to the northeast. Also note the circles in Figure 2 denoting five representative positions in the OH data discussed in the next paragraph. These five circles are also shown in the finder chart of Figure 1b (larger circles). The four OH absorption regions are as follows: (1) The most prominent absorption is toward the Northeast Dark Lane along the northeast edge of M42. Here, peak 1667 MHz optical depths reach $\tau_{0,1667} \approx 1$, perhaps higher in regions of saturation. (2) Much weaker OH absorption exists toward the Dark Bay. This absorption appears in Figure 2 as a tongue extending westward from the Northeast Dark Lane toward the brightest region of radio continuum emission. OH optical depths are highest ($\tau_{0,1667} \approx 0.04$) in the direction of Knot 1 identified by O'Dell & Yusef-Zadeh (2000) in their optical extinction image. (3) OH absorption ($\tau_{0,1667} \approx 0.12$) is observed approximately one arc min to the southwest of the peak continuum brightness. This direction closely coincides with the Orion-S molecular core (e.g. Henney et al. 2007), and the absorption appears as a distinct, nearly circular region, suggesting it is barely resolved in the VLA data. We refer to this absorption as the Orion-S OH absorption, although the connection between OH and Orion-S bears further consideration. See section 4.4. The Orion-S OH absorption feature has a clear counterpart in H I absorption, with peak optical depth $\tau_{0,HI} \approx 0.2$ in the present data and 0.45 in the higher spatial resolution data of vdWGO13. See section 4.4 and Figure 10. Finally, (4) strong OH absorption ($\tau_{0,1667} \approx 0.6$) appears along the eastern edge of M43. This absorption is clearly associated with the dark filament of optically obscuring material in Figure 1a. This same filament is visible in the 800 µm OMC-1 dust continuum image of Johnstone & Bally (1999).

To illustrate the nature of OH absorption toward Orion, we choose five representative positions in our image data. We label these positions OH1 – OH5. Two of these positions are located in the Northeast Dark Lane, and one is located in each of the Dark Bay, Orion-S, and M43 regions. The locations of these five positions are shown in Figures 1a, 1b and 2 as circles of 40" diameter, the 1665 and 1667 MHz synthesized beamwidth. Table 2 lists these positions and their relevant parameters; Table 1b lists the channel separation and resolution. Figure 3 shows 1667 MHz OH optical depth profiles for each position. Note that the optical depth profile for Orion-S is about a factor of two wider than any of the other profiles. Comparison of the OH optical depth profiles with H I optical depth profiles at these five positions (the latter not shown) indicates that the OH lines always lie within the velocity range of the saturated H I lines. Except for the Orion-S position (Figure 10), it is not possible to identify OH lines with specific H I velocity components owing to H I line saturation.

For each of the five representative positions of OH absorption (Table 2), we derive $N$(OH) from the velocity integrated 1665 MHz and 1667 MHz optical depth profiles. To do so, we use the standard conversion factors $C_{1665}$ and $C_{1667}$ from Roberts et al. (1995, see above), and we assume $T_{ex}$ = 20 K for both OH lines. However, N(OH) ∝ $T_{ex}$, and $T_{ex}$ cannot be measured from our absorption line data. $T_{ex}$ can only be measured from a combination of absorption and emission lines along the same line of sight. Using such galactic data toward extragalactic continuum sources, Liszt & Lucas (1996) found $T_{ex}$ ≈ 4 – 13 K for the OH main line transitions. Recently, Heiles (private communication) analyzed a similar but more extensive data set from the Millennium Survey (Heiles & Troland, 2005). Heiles found $T_{ex}$ for the OH main lines in the range 2-15 K, with the majority of the values below 10 K and many less than 5 K. These OH data toward extragalactic continuum sources likely sample low density molecular gas ($n$(H) ≈ $10^{2-3}$ cm$^{-3}$) where the OH lines are very subthermally excited ($T_{ex}$ << $T_K$). OH bearing gas in the Veil is likely to be much denser (see note *a* to Table 4b) and somewhat warmer, the latter owing to its proximity to the Trapezium. Therefore, we expect $T_{ex}$ to be higher in the Veil than in the low density molecular gas. According to the theoretical study of Guibert et al. (1978), $T_{ex}$ for the OH main lines is rather insensitive to $T_K$. However, $T_{ex}$ is sensitive to $n$(H). For a gas with $T_K$ = 50 K, Guibert et al. calculate that $T_{ex}$ increases from 5 K to an asymptotic value of 30 K in the range $n$(H) ≈ $10^{2-4}$ cm$^{-3}$ (see their Figure 2). Given this trend, we believe that our assumption of $T_{ex}$ = 20 K for the Veil OH lines is reasonable and likely to be correct to within a factor of two.

Having estimated N(OH) as described above, we use standard conversion factors to scale $N$(OH) to $N$(H) and to $A$v, where $N$(H) = $N$(H$^0$) + 2 $N$(H$_2$). Results are given in Table 2 along with the conversion factors we have used. Note that values of $A$v in Table 2 are *not* direct measures of optical extinction like those reported by O'Dell & Yusef-Zadeh (2000). Judging from Table 2, most of the OH absorption arises in high column density regions ($A$v > 50). The only exception is the Dark Bay for which $A$v ≈ 7. Also, we note that the ratios of peak OH optical depths $\tau_{0,1667}/\tau_{0,1665}$ deviate from the LTE value of 1.8 for several of the positions in Table 2. Ratios greater than the LTE value indicate non-thermal excitation of the 18 cm OH levels, a common phenomenon (e.g. Crutcher 1977). Ratios less than the LTE value may also indicate non-thermal excitation and/or clumping of optically thick gas on scales smaller than the synthesized beamwidth.

### 3.4. The H I Zeeman Effect

A principal goal of these observations is to study the distribution of magnetic field strengths in the Veil. The strong radio continuum emission of the Orion H$^+$ region leads to strong H I absorption lines, and, via the H I Zeeman effect, to measurements of $B_{los}$ in the H$^0$ layers of the Veil. Aperture synthesis data yield images of $B_{los}$ across the Veil, and they do so independently for different velocity components. For the Veil data, we found evidence of the Zeeman effect in both velocity components A and B. Accordingly, we fitted separately over the channel ranges appropriate to component A (5.3 to 22 km s$^{-1}$) and component B (-14.7 to 3.6 km s$^{-1}$). Note that these velocity ranges include line-free channels necessary for the least squares fits. We discuss below the images of $B_{los}$ derived for these two components. We note that sensitivity to $B_{los}$ is highly variable among pixels in the images. This variability arises from variations across the Veil in radio continuum brightness, line optical depth, and line width. Higher sensitivities to $B_{los}$ arise in pixels where the radio continuum brightness and line optical depths are higher and where lines are narrower. Velocity component A has generally higher optical depths and narrower lines than component B. Therefore, the image of $B_{los}$ for component A has better sensitivity over a larger number of pixels. In Figure 4, we show an example of 21 cm Stokes I/2 and V/2 profiles. These profiles are for a pixel in the images lying close to the center of the Trapezium. At this particular position, $B_{los}$ is the same to about 1-σ for the two velocity components. Table 1a lists the channel separation and resolution. The synthesized beamwidth is 25".

### 3.4.1. The H I Zeeman Effect in Component A

In Figure 5 we present an image of $B_{los}$ in Veil H I velocity component A. Figure 5 (synthesized beamwidth 25") is one of the most detailed images of interstellar magnetic field strengths published to date. $B_{los}$ in component A is uniformly *negative* (magnetic field pointed *toward* the observer). For simplicity, we use the symbol $B_{los}$ below to signify $|B_{los}|$. Also, Figure 5 has been masked to include only those pixels for which $|B_{los}|/\sigma(B_{los}) > 3$. Perhaps the most striking characteristic of Figure 5 is the relative uniformity of $B_{los}$ over much of the Veil, with most values in the range 30 to 60 µG. However, there is a tendency for $B_{los}$ to increase gradually from southwest to northeast, roughly similar to the increase in H I optical depths of components A and B. (See vdWG89.) Also, there appears to be a ridge of enhanced $B_{los}$ extending from the southwest diagonally across the Veil to the northeast and widening toward the northeast. In this ridge, $B_{los} \approx 50$ to 60 µG, whereas on either side of the ridge $B_{los} \approx 30$ to 40 µG. The ridge of enhanced $B_{los}$ coincides approximately with a ridge of enhanced $N(H^0)/T_{ex}$ in the image of component A presented by vdWG89. An indication of the ridge also appears in the H I optical depth image of vdWGO13 for a velocity of 8.4 km s$^{-1}$, on the unsaturated red wing of H I component A. Finally, there are isolated regions of higher $B_{los}$ in Figure 5, most obviously in the northeast sector of the source. Here, $B_{los}$ is typically > 100 µG, reaching 300 µG in one small region about 30" in size. Other isolated regions of higher $B_{los}$ (75-100 µG) lie in the northern part of the Veil and near the southwest edges of the image in Figure 5. Overall, the image of $B_{los}$ for component A is consistent with the equivalent image in Troland et al. (1989) where component A is labeled the high-velocity (HV) component. However, the present results in Figure 5, with their higher sensitivity and spatial resolution (25" vs. 40"), include values of $B_{los}$ for more independent pixels, and they reveal high field strengths ($B_{los} > 100$ µG) in the northeast sector of the Veil, a region not included in the earlier results of Troland et al.

### 3.4.2. The H I Zeeman Effect in Component B

In Figure 6 we present an image of $B_{los}$ in Veil H I velocity component B (synthesized beamwidth 25"). Again, we have masked the figure to include only pixels for which $|B_{los}|/\sigma(B_{los}) > 3$. As for component A, all values are *negative*, and we use the symbol $B_{los}$ below to signify $|B_{los}|$. Since component B is generally weaker and wider than component A, sensitivity to $B_{los}$ is lower, and Figure 6 displays fewer values of $B_{los}$ than Figure 5. Overall, the image of $B_{los}$ for component B is remarkably similar to that for component A. In particular, $B_{los}$ for component B is quite uniform over most of the image, with most values $B_{los} \approx 50$ to 60 µG. Also, values for $B_{los}$ in component B increase toward the northeast sector of the source, reaching 100 µG or more. This behavior, too, is similar to that for component A. In detail, however, there are differences in the distribution of $B_{los}$ for the two H I velocity components. For example, a peak in $B_{los}$ for component B in the southwest region of the image (90 µG) nearly coincides in position with a minimum in $B_{los}$ for component A (30 µG).  The image of $B_{los}$ for component B is consistent with the equivalent image in Troland et al. (1989) where component B is labeled the low velocity (LV) component.  However, the earlier image of $B_{los}$ for this component includes a very few independent measurements owing to severe sensitivity limitations.

### 3.5. The OH Zeeman Effect

The present data provide the first aperture synthesis measurement of $B_{los}$ in molecular regions of the Veil.  Previously, Troland et al. (1986) detected of the 1667 MHz OH Zeeman effect towards Orion A using the Nancay telescope.  They found $B_{los,1667}$ = -125 ± 20 µG, positioning the Nancay 3.5' × 19' beam (FWHM, right ascension and declination, respectively) 3' east of the Orion A radio continuum peak (see contours in Figure 2).  As for the HI data (section 3.4), the OH data have highly variable sensitivities to the Zeeman effect.  Careful examination of our 1667 MHz Stokes I and V profiles reveals the Zeeman effect reliably detected in only one independent position, OH1 in the Northeast Dark Lane (Table 2).  Here, $B_{los,1667}$ = -350 ± 47 µG and $B_{los,1665}$ = -151 ± 46 µG.  At this same position, $B_{los,H\,I}$ is -183 ± 32 µG in H I component A.  In Figure 7 we present Stokes I/2 and V/2 profiles for the 1665 and 1667 MHz lines.  The synthesized beamwidth is 40"; the channel separation is 0.27 km s$^{-1}$.  The center velocity of the lines in this figure coincides quite well with the center velocity of the lines observed by Troland et al. (1986) at Nancay.  Evidently, these authors sampled with a single dish nearly the same region of the Veil as we did with the VLA.  The higher spatial resolution of the VLA likely accounts for the higher field strength detected with the aperture synthesis instrument.

Values of $B_{los}$ derived from the 1665 and 1667 MHz lines only agree to within 2-σ.  Also, the 1665 MHz Stokes V/2 profile in Figure 7 shows scant signs of the Zeeman effect.  The Orion field of view includes OH masers that we attempted to remove in the uv-plane **(Section 2).** These masers are much stronger in the 1665 MHz line, and their effect cannot be completely removed from the synthesis data.  Therefore, we suspect that residual maser contamination is present in the 1665 MHz Stokes V/2 profile of Figure 7.  Troland et al. (1986) noted the same phenomenon.  Their single-dish 1665 MHz Stokes V profile shows clear evidence for maser contamination near the thermal OH absorption line velocity.  Yet they find no such contamination in their 1667 MHz Stokes V profile.  With these considerations in mind, we assume that our Zeeman effect result at 1667 MHz is more reliable than that at 1665 MHz, so we use the former result in our energy analysis of section 4.2.

### 4. DISCUSSION

### 4.1. Morphological Connections Among $B_{los}$, $N(H^0)$ and Optical Extinction

The present data allow for a comparison among images of different physical parameters across the Veil.  These images include $B_{los}$ in H I components A and B, $N(H^0)/T_{ex}$ in components A and B, and optical extinction across the Veil. (For orientation purposes, recall that the dashed yellow rectangle in the optical image of Figure 1a corresponds to the fields of view of the $B_{los}$ maps in Figures 5 and 6.)  Comparisons among images are necessarily incomplete because images of the relevant parameters are incomplete over different parts of the Veil.  For example, images of $N(H^0)/T_{ex}$ only provide information in the southwest regions of the Veil where the H I line is unsaturated.  Images of $B_{los}$ are limited by sensitivity to the Zeeman effect, they only cover the inner parts of the Veil (especially for H I component B), and the images are only sensitive to one component of the magnetic field. The optical extinction images, likewise, only cover the inner Veil where O'Dell & Yusef-Zadeh (2000) could measure extinctions.

Despite the limitations in image coverage, we find morphological similarities between the distributions of $N(H^0)/T_{ex}$ and magnetic fields across the Veil. In particular, there is a significant and previously established (vdWG89) tendency for the velocity-integrated $N(H^0)/T_{ex}$ to increase from the southwest toward the northeast of the Veil. This trend also applies individually to H I components A and B. Likewise, there is a significant tendency for $B_{los}$ in H I component A to increase in the same direction, becoming highest in the direction of the Northeast Dark Lane where the H I line is deeply saturated (Figure 5, section 3.4). The image of $B_{los}$ in H I component B (Figure 6) is more limited in coverage, so this same trend is not so obvious. Nonetheless, the highest values of $B_{los}$ for component B also occur in the northeast region of the image, just as they do for component A. Considering the morphology in slightly more detail, both $N(H^0)/T_{ex}$ in H I component A (vdWG89) and $B_{los}$ in the same component display a ridge of higher values along a southwest to northeast axis. These morphological similarities between H I and $B_{los}$ suggest a general tendency toward constant mass-to-flux ratio ($\propto N(H^0)/B_{los}$) in the Veil. The similarities in morphology among $N(H^0)/T_{ex}$ in components A and B and $B_{los}$ in components A and B imply a very close association between these two components despite their difference in center velocities.

The morphology of the optical extinction image (O'Dell & Yusef-Zadeh 2000) bears a more complex relationship to the morphologies of the $N(H^0)/T_{ex}$ and $B_{los}$ images. As a result, it seems unlikely that the Dark Bay is directly associated with H I components A and B. Optical extinction increases from southwest to northeast across the image. Also, there is some indication of a ridge of enhanced extinction along this axis. This morphology is similar to those of $B_{los}$ and $N(H^0)/T_{ex}$ in components A & B (see above). Indeed, the morphological similarities between optical extinction and $N(H^0)/T_{ex}$ across the Veil led O'Dell & Yusef-Zadeh (2000) to conclude that most of the optical extinction of the Orion Nebula arises in the $H^0$ layers of the Veil rather than in the ionized gas itself. However, there is one obvious exception to these morphological similarities, the Dark Bay. No indication of the Dark Bay appears in the images of $B_{los}$ (Figures 5 & 6). Therefore, it seems likely that the Dark Bay is physically separated from Veil H I components A and B. This suggestion is consistent with the morphology of optically obscuring material shown qualitatively in Figure 1a. In this optical image, the Dark Bay appears to be a westward extension of very optically obscuring material to the east and northeast of the Orion Nebula. The optically obscuring material includes not only the Dark Bay. It also includes the Northeast Dark Lane and the filament of obscuring material partially covering M43. Perhaps this whole complex of optically obscuring material is a part of OMC-1 that lies in front of Veil H I components A and B.

Finally, we comment upon the connection between OH absorption and the optical extinction image. OH absorption is only observed in four regions of the Veil (section 3.3). Of these four regions, only two lie within the area covered by the optical extinction image of O'Dell & Yousef-Zadeh (2000). These regions are the Dark Bay and Orion-S, where Av in the optical extinction image is about 5.5 and 1.4 mag, respectively. There is no detectable OH absorption toward the Trapezium stars ($A$v $\approx$ 2). Also, there is no OH absorption toward the Southwest Cloud (SW Cloud in O'Dell & Yusef-Zadeh, 2000) about 2' southwest of the Trapezium where $A$v $\approx$ 3. The Southwest Cloud appears in Figure 1a as a distinct, scorpion-shaped patch of extinction near the southwest edge of the bright $H^+$ region. If we exclude Orion-S from consideration, as a different environment (see Section 4.4), then OH absorption is only observed in the Veil when Av > 5. The absence of OH absorption except at relatively high Av is not surprising since OH requires significant shielding to survive in an environment of such intense stellar radiation. This statement is consistent with the Cloudy spectral synthesis code model of Orion-S (O'Dell et al. 2009). In this model, significant OH does not form until $A$v reaches about 4 mag. Although physical conditions in the Orion-S core differ significantly from those in the Veil, the Cloudy model incorporates the same intense stellar radiation field that is incident upon the Veil. Therefore, the Cloudy predictions about OH formation as a function of $A$v are likely to be relevant to the Veil in general, not just Orion-S.

### 4.2. Energetics in the Orion Veil and elsewhere

The present observations offer insights into thermal, turbulent, magnetic and gravitational energies in the Veil. Abel et al. (2006) performed an energy analysis for the single Veil line of sight toward the Trapezium stars. (See their Table 5.) Here we expand the analysis to many more lines-of-sight through the Veil, and we compare Veil energetics with those of other atomic and molecular regions in the ISM.

The energetics of the Veil and elsewhere can be characterized by the thermal, turbulent and magnetic energy densities (erg cm$^{-3}$), where $E_{therm} = (3/2) nkT_K$, $E_{turb} = (3/2) \rho\sigma_{turb}^2$, and $E_B = B^2/8\pi$. In these relationships, $n$ is the particle density (cm$^{-3}$) including He, $\rho$ is the mass density (g cm$^{-3}$) including He, $\sigma_{turb}$ is the 1-D turbulent velocity *dispersion*, and $B$ is the *total* magnetic field strength ($B_{tot}$), not the measured $B_{los}$. Note that for velocity dispersions, $\sigma = \Delta V_{FWHM} / (8 \ln 2)^{1/2}$, where $\Delta V_{FWHM}$ is the velocity full width at half maximum. Several other parameters relate to the *ratios* of energy densities. See the discussion in Heiles & Troland (2005) and in Crutcher (1999). These parameters include the *turbulent Mach number* $M_{turb}$, where $(1/3) M_{turb}^2 = E_{turb}/E_{therm}$, assuming isotropic turbulence. The *conventional plasma parameter* $\beta_{therm} = (2/3) E_{therm}/E_B$. Note that $\beta_{therm}$ is sometimes defined to be one half this value (Gammie & Ostriker, 1996). The *turbulent plasma parameter* $\beta_{turb} = (2/3) E_{turb}/E_B$. Therefore, $\beta_{turb} = M_A^2$, where $M_A$ is the *Alfvenic turbulent Mach number*. Finally, the dimensionless *mass-to-flux ratio* $\lambda$ is a measure of the ratio of gravitational-to-magnetic energies. If $\lambda < 1$, magnetic energy dominates gravity, and the material is magnetically *subcritical*. If $\lambda > 1$, gravity dominates magnetic energy, and the material is magnetically *supercritical*. We take $\lambda = 5\times10^{-21} N(H)/B$ where $B$ is in µG (Crutcher 1999). Note that $\beta_{therm}$ and $\beta_{turb}$ are defined as ratios of *pressures* rather than energy densities. In the discussion above, we take $E_{therm} = (3/2) P_{therm}$; $E_{turb} = (3/2) P_{turb}$; and $E_B = P_B$, where $P_{therm}$, etc. are the associated pressures.

Several difficulties arise in performing an energy analysis for the Veil. For one, relevant physical parameters such as $N(H^0)/T_{ex}$ and $B_{los}$ are not uniformly sampled in the images (section 4.1). Therefore, we have adopted a semi-quantitative approach in which we choose nine representative positions in the images to evaluate energy parameters in the H$^0$ gas. These positions are labeled HI1 through HI9; they are shown in Figure 1a (blue circles), Figure 1b (smaller black circles), and Figure 5 (blue circles). A subset of these positions is shown in Figure 6. All positions lie in the southwestern region of the Veil where the H I line is unsaturated so $N(H^0)/T_{ex}$ is known. Coordinates and measured parameters for these positions are given in Table 3. We also choose five positions in the images to evaluate energy parameters in the molecular gas via OH absorption. These positions are listed in Table 2 and shown in Figure 1a (yellow circles), Figure 1b (larger black circles) and in Figure 2 (black circles).

A second difficulty in performing an energy analysis for the Veil lies in the uncertain systematic errors in many of the derived energy parameters. Energy parameters are calculated from direct measurements of line widths $\Delta V$ as well as estimates of the quantities $T_{ex}$, $T_K$, $N(H)$, $n(H)$, and $B_{tot}$. Uncertainties exist in each of these parameters. For example, estimates of $B_{tot}$ are based upon statistical corrections to the measured values of $B_{los}$. (See notes to Tables 4a and 4b for the statistical corrections we applied to estimate $B_{tot}$ from $B_{los}$.) Values of $n(H)$ are based upon PDR models of the Veil (Abel et al. 2006, 2016) or else upon simple geometrical assumptions, as specified in notes to Tables 4a and 4b. Estimates of $N(H)$ in the molecular gas depend upon an assumed value of $T_{ex}$ (section 3.3) and of the ratio $N(OH)/N(H)$. No rigorous technique exists to quantify the propagation of systematic errors into the calculation of energies in the ISM. Therefore, conclusions drawn from these calculations must be regarded as approximate guides to the energetics of the region, accurate, perhaps, to a factor of two at best.

Results of our energy analysis are listed in Tables 4a and 4b. In Table 4a we list values of energy parameters in H I components A and B. For comparison purposes we also list energy parameters for the Cold Neutral Medium (CNM) given by Heiles & Troland (2005). In Table 4b we list equivalent information for molecular gas in the Veil. For comparison purposes, we provide similar information for an ensemble of low mass molecular cores and for molecular cores associated with three regions of massive star formation, M17, S106 and S88B. Information about magnetic field strengths in low mass molecular cores and in the three regions of massive star formation comes from the literature of 1665 and 1667 MHz OH Zeeman effect observations. See notes to Table 4b for references. Note that Orion-S is included with the Veil in Table 4b. However, this high-mass molecular core is likely to be quite different from the other Veil regions, and it is unlikely to be part of the Veil at all (section 4.4). In both Tables 4a and 4b we also list the *per cent* contributions of $E_{therm}$, $E_{turb}$ and $E_B$. See notes for these tables. With due regard for uncertainties, a variety of conclusions can be drawn from the energy parameters of Tables 4a and 4b. We list these conclusions below, many of which are consistent with conclusions previously drawn for other regions of the neutral ISM:

(a) *Turbulent and thermal energies ($M_{turb}$)* - Turbulent energy is greater than thermal energy in the atomic and molecular regions of the Veil (supersonic turbulence, $M_{turb} > 1$). H I components A and B are only mildly supersonic with $M_{turb} \approx 3$ ($E_{turb}/E_{therm} \approx 3$). In this sense, they are very similar to the CNM although they are of order one dex denser. The molecular regions of the Veil are much more highly supersonic, with $M_{turb} \approx 4\text{-}6$ ($E_{turb}/E_{therm} \approx 5\text{-}12$). In this sense, the Veil molecular regions are very similar to molecular cores in other high mass star forming regions included in Table 4b. High mass star formation is associated with very turbulent molecular gas. However, low-mass molecular cores associated with low-mass star formation (Table 4b) fall into a different category. These regions, when sampled in 1665 and 1667 MHz OH lines, are close to transonic, with $M_{turb} \approx 2$ ($E_{turb}/E_{therm} \approx 1$). As numerous previous studies have shown, the neutral ISM is a very turbulent environment in which thermal energy is relatively insignificant. An exception to this rule is the low-mass molecular core.

(b) *Thermal and magnetic energies ($\beta_{therm}$)* – Thermal energy is insignificant compared to magnetic energy in HI components A and B and in the molecular Northeast Dark Lane 1 position ($\beta_{therm}$, $E_{therm}/E_B << 1$). The same statement applies to the three molecular cores associated with high mass star formation listed in Table 4b. Thermal energy is (at least on the average) more significant in the CNM ($\beta_{therm} \approx 0.3$, $E_{therm}/E_B \approx 0.5$) and, especially, in low mass molecular cores ($\beta_{therm} \approx 1.3$, $E_{therm}/E_B \approx 2$).

(c) *Turbulent and magnetic energies* ($\beta_{turb}$, $M_A$) – A rough equipartition exists between turbulent and magnetic energies in most regions included in Tables 4a and 4b (trans-Alfvenic turbulence, $\beta_{turb} \approx 1$, so $M_A \approx 1$). Such a balance has been previously suggested for regions of the ISM sampled by the Zeeman effect (e.g. Myers & Goodman, 1988, Crutcher 1999). In this sense, the atomic and molecular gas of the Veil is similar to atomic and molecular gas elsewhere in the ISM. However, there is one very notable exception to the equipartition rule, H I component A. Here, $\beta_{turb} \approx 0.02$, so $M_A \approx 0.1$ and $E_{turb}/E_B \approx 0.03$. That is, this layer is very strongly dominated by magnetic energy over turbulence, *i.e.* very sub-Alfvenic turbulence. A similar conclusion was drawn by Abel et al. (2006) for the line-of-sight to the Trapezium alone. In section 4.3.2 we comment further upon the possible origin of sub-Alfvenic turbulence in HI component A. We note that Zeeman effect studies have yet to identify any examples of super-Alfvenic turbulence ($\beta_{turb}$, $M_A \gg 1$) in the ISM. Evidently, super-Alfvenic turbulence is damped on short timescales.

(d) *The balance of thermal, turbulent and magnetic energies* – Tables 4a and 4b list the *per cent* contributions of these three energies to regions in the Veil and elsewhere. In general, magnetic and turbulent energies are comparable, with thermal energy relatively small. Obvious exceptions to this statement arise in (1) HI component A, where magnetic energy dominates the other two, and in (2) low mass molecular cores, where, on average, thermal energy is significant and comparable to the other two. The CNM is also a region where thermal energy is not negligible, although turbulent and magnetic energies are greater.

(e) *Thermal, turbulent and magnetic energies combined (total pressure)* – The sum of these three energy densities is a measure of the total support of a cloud or core against the confining effects of gravitation and external pressure. These sums are listed in the bottom rows of Tables 4a and 4b. (For thermal and turbulent motions, pressures are 2/3 times energy densities.) Examination of Tables 4a and 4b leads to the identification for four distinct energy density or pressure regimes, with values ranging over four dex. These regimes, in order of increasing total energy density, are: (1) the CNM; (2) low mass molecular cores, one dex higher; (3) H I components A and B, one dex higher; and (4) molecular gas associated with high mass star formation (Orion Veil, M17, S106, S88B), two dex higher. Note that the total energy densities in H I components A and B are approximately the same despite the different distribution among the energies. In effect, these two $H^0$ regions are in approximate pressure equilibrium. Note, also, that the total energy density of the CNM (Table 4a) is very comparable to the galactic midplane pressure of $3.9 \pm 0.6 \times 10^{-12}$ dyn cm$^{-2}$ derived by Boulares & Cox (1990). That is, the CNM pressure is in equilibrium with the weight in the z-direction of the midplane gas.

(f) *Magnetic and gravitational energies ($\lambda$)* – As listed in Table 4a, $\lambda < 1$ in both HI components A and B (*i.e.* $H^0$ is subcritical), and the value of $\lambda$ is very similar to the average value for the CNM ($\lambda \approx 0.1$). That is, magnetic energies dominate gravitational energies in both Veil $H^0$ regions as they do in the CNM. However, $\lambda \approx 1$ in the molecular Northeast Dark Lane 1 position where $N(H)$ is about 2 dex higher than in HI components A and B (Table 4b). This distinction between low-$N(H)$ regions that are *subcritical* ($\lambda < 1$) and higher-$N(H)$ regions that are *critical to supercritical* ($\lambda \geq 1$) is consistent with a large body of Zeeman effect results covering $N(H) \approx 10^{19-24}$ cm$^{-2}$ as described below.

    To illustrate the relationship between $\lambda$ and $N(H)$, we present existing Zeeman effect data in Figure 8 on the log $N(H)$ – log $B_{los}$ plane. Veil data points are shown as solid colored circles; data for H I component A are in red, data for H I component B are in blue, and the data point for

OH is in green. Other data points (black circles) were plotted by Crutcher (2012) who lists relevant references, or else they are from Thomson, Troland & Heiles (in preparation). For $N(H) \leq 10^{21}$ cm$^{-2}$, the black circles sample diffuse H$^0$ gas via the 21 cm HI line. For $N(H) \approx 10^{21-22}$ cm$^{-2}$, the black circles represent molecular gas sampled by 18 cm (1665, 1667 MHz) OH lines. For $N(H) > 10^{23}$ cm$^{-2}$, the black circles denote dense molecular cores sampled by the 3 mm CN lines. (The $N(H)$ ranges are approximate.) The diagonal line in Figure 8 represents $\lambda = 1$. Points above the line are magnetically subcritical (magnetically dominated, $\lambda < 1$); points below the line are magnetically supercritical (gravity dominated, $\lambda > 1$). Note that values of $B_{los}$ are plotted directly without statistical correction. Therefore, a given point in Figure 8 represents an *upper limit* upon the true value of $\lambda \propto N(H)/B_{tot}$ since a measured value of $B_{los}$ represents a lower limit upon $B_{tot}$. Given an ensemble of $B_{los}$ measurements, as shown in Figure 8, the upper envelope is an indication of $B_{tot}$, as a function of $N(H)$, assuming a random distribution of field angles relative to the line-of-sight.

The Zeeman effect data of Figure 8 and of Tables 4a and 4b illustrate several key points. First, we exclude from consideration the colored points for the Veil. Then, magnetic field strengths (black circles) remain remarkably constant with increasing $N(H)$ up to $N(H) \approx 1\text{-}2 \times 10^{22}$ cm$^{-2}$. For higher N(H), field strengths increase. Otherwise stated, $\lambda$ increases systematically with increasing $N(H)$, remaining subcritical at lower values of $N(H)$ in the diffuse H$^0$ gas. Above $N(H) \approx 1\text{-}2 \times 10^{22}$ cm$^{-2}$, $\lambda$ in the molecular gas becomes nearly constant and slightly supercritical. The constancy of field strengths at lower $N(H)$ has been noted in the past (e.g. Crutcher 2012). Lazarian et al. (2012) suggest it is the result of the reconnection diffusion process that maintains approximately constant field strength in a gathering cloud until the free fall time scale becomes shorter than the reconnection diffusion time scale. Once this threshold is reached, the gas contracts gravitationally, drawing in and strengthening the field in the process. We comment further on this process in section 4.3.2.

Now considering the colored points in Figure 8 for the Veil, the green circle representing molecular gas in the Northeast Dark Lane is quite consistent with other data points for molecular gas of comparable N(H). Very likely, this region of the Veil has been largely undisturbed by the star formation process, so it is typical of dense molecular gas associated with massive star formation elsewhere. However, the red and blue circles representing HI components A and B, respectively, lie a factor of 3-4 above the black circles for magnetic fields in other regions of the ISM with comparable N(H). Otherwise stated, these Veil H$^0$ regions are more magnetically subcritical (i.e. $\lambda$ smaller) than other regions of comparable N(H). See section 4.3 for further comments about the low values of $\lambda$ (i.e. high ratios B/N(H)) in these components.

### *4.3. The natures of H I Velocity components A & B*

The Orion Veil offers an unusual opportunity to explore PDR physics, as noted in the Introduction. The high field strengths in the Veil compared to those in the CNM, as well as a wealth of other observational evidence (*e.g.* vdWGO13), clearly establish that the Veil is closely associated with the Orion star forming region, not unrelated diffuse gas along the line of sight. Also, the Veil has apparent plane parallel geometry perpendicular to the line-of-sight, and it is located nearby, so it is susceptible to high spatial resolution studies. Abel et al. (2016) constructed detailed PDR models of the Veil. (See, also, Abel et al. 2004 and 2006.) From these models, they estimate $n(H) \approx 10^{2.5}$ and $10^{3.4}$ cm$^{-3}$ and $T_K \approx 50$ and 60 K for components A and B, respectively. Strictly speaking, these parameters apply only to the line of sight toward the Trapezium. However, we assume they apply to other Veil lines of sight where N(H) is comparable.

The magnetic field images in Figures 5 & 6 offer important additional perspectives on the natures of H I components A and B. These images reveal that $B_{los}$ in the two H I components have very similar values and morphologies (section 4.1). These magnetic similarities, together with the similarities in morphology of $N(H^0)/T_{ex}$ and in velocity gradients between the two components (see vdWG89), establish that components A and B are very closely associated. However, two questions must be addressed before a comprehensive understanding of the Veil is possible. One question is the distances of components A and B from the principal source of radiation, the Trapezium stars. The second question is the origin of the relatively high magnetic field strengths in components A and B. We consider these questions in the following two subsections.

*4.3.1. Locations and Natures of H I Components A and B*

Several authors have concluded that H I component B lies closer to the Trapezium stars than component A. For example, vdWG89 argued that component B consists of gas heated and photodissociated by non H-ionizing stellar photons escaping from the $H^+$ region. In this picture, H I component A is gas that has not yet been kinematically affected by the $H^+$ region and makes up an envelope of $H^0$ surrounding the molecular gas. These authors also noted that the blueshift of component B relative to A ($\approx 4$ km s$^{-1}$; see section 3.2) can result from thermal expansion of the former owing to radiative heating. Recently, Abel at al. (2016) obtained high spectral resolution Hubble Space Telescope (STIS) spectra in the 1133 – 1335 nm wavelength range toward Theta $^1$B Ori. These spectra resolve HI components A and B. Abel et al. detected Veil absorption lines of CI, CI*, CI** and rotationally/vibrationally excited $H_2$. Based on these data, they constructed a PDR model for the Veil that yields estimates of the distances of components A and B from the Trapezium as well as the temperatures, volume densities and thicknesses of the two components. The model establishes that component B is, indeed, closer to the Trapezium, about 2 pc distant, as compared to 2.4 pc for component A.[2] This relative placement is consistent with the facts that component B is warmer, thinner, denser and more turbulent than component A. All of these properties point to more interactions of component B with the Trapezium environment than component A. At the same time, kinematic and magnetic similarities suggest a very close physical association between components A and B, as argued above.

One possible close physical association between the two components is that they lie on opposite sides of the shock preceding a weak-D ionization front. That is, component B is shocked gas, and component A is pre-shock gas. For this scenario to apply, we assume that the magnetic field lies nearly perpendicular to the plane of the shock (i.e. along the line-of-sight). Then, neglecting possible field tangling induced by motions within the shock layer itself, the shocked gas would slide along the field lines, increasing its density with no increase in magnetic field strength in component B, as observed. In such a case, standard, non-magnetic jump conditions apply for the shock since the magnetic pressure is not relevant to motions along the field direction. We apply the equations for non-magnetic shocks in Shu (1992), using values for $n$(H) and $T_K$ in components A and B derived by Abel et al. (2016; see section 4.3 above). We estimate a shock velocity of 2 km s$^{-1}$. The shock results in a blueshift of component B relative to component A of 1.7 km s$^{-1}$, to be compared with the observed value of 4 km s$^{-1}$. That is, the shock model underpredicts the velocity difference between the two components by a factor of about two. The predicted velocity difference is approximately proportional to $[T_B \times n(H)_B/n(H)_A]^{1/2}$, where $T_B$ is the kinetic temperature of component B. Therefore, a two times higher predicted velocity difference would require, for example, a two times higher temperature *and* two times higher density ratio. Abel et al. (2016) do not provide error estimates for their photoionization model-derived values of $T_B$, $n(H)_B$, and $n(H)_A$. As a result, we cannot estimate the likelihood of higher temperatures and density ratios. Also, the velocity difference between components A and B varies as a function of position across the Veil, with values closer to 3 km s$^{-1}$ in some positions. In short, the shock model is attractive because it accounts for kinematic and magnetic similarities between the components, and the shock passage may account for the higher turbulence in component B than in A. However, the shock model offers an uncertain quantitative fit to photo-ionization model densities and temperatures of Abel et al. (2016).

---

[2] van der Werf et al. (2013) argue that Component B is about 0.4 pc from the Trapezium. However, Abel et al. (2016) present compelling evidence that the distance is about 2 pc.

We illustrate the natures and locations of Veil components A and B with the cartoon of Figure 9. In this figure, the Earth is to the left, and north is up. Neutral gas is shown in blue, ionized gas, in red. The cartoon places component B closer to the Trapezium, and the cartoon shows the thicknesses of components A and B as derived from the Abel et al. (2016) Veil model, 1.6 pc and 0.40 pc, respectively. According to this same model, $n(H) \approx 10^{2.5}$ cm$^{-3}$ and $10^{3.4}$ cm$^{-3}$ for components A and B, respectively. The higher density of component B is suggested in Figure 9 by darker shading. Three H$^+$ regions are shown in the figure. One is the main Orion H$^+$ region (Orion Nebula) with a scale height of 0.05 pc (O'Dell 2001), located 0.25 pc behind the Trapezium stars (Abel et al. 2006). The vertical extent of the Orion Nebula in the cartoon represents the size of the bright Huygens region, about 0.8 pc. (The vertical extent of the Veil is arbitrary.) The second H$^+$ region in Figure 9 is labeled "Blue-shifted H$^+$" ("blue component" in Abel et al. 2016; also described by Abel et al. 2006). This matter-bounded, ionized layer is shifted about -25 km/s relative to OMC-1, and it is observed in optical emission lines and uv absorption lines. Abel et al. (2006) note that the blue shift can be ascribed to stellar radiative acceleration over a period of order 10$^5$ years. The third H$^+$ region, labeled "Veil H$^+$ region", is presumed to exist on the front face of the neutral Veil. However, optical and uv tracers of this layer would lie at velocities similar to those of the Orion Nebula, hence, they are unobservable. Immediately behind the Orion Nebula lies the OMC-1 filamentary cloud and the Orion-KL core. Also shown is the Orion-S molecular core embedded within OMC-1. However, this Orion-S location is only one possibility discussed in section 4.4 below.

Finally, we comment upon the 21 cm HI excitation or "spin" temperatures $T_s$ in components A and B in comparison with their kinetic temperatures. Under typical conditions in the galactic ISM, it is commonly expected that $T_s \approx T_K$. However, under conditions of low density and/or high stellar uv flux, resonant scattering of Lyman-alpha photons can pump the HI levels, increasing $T_s$ above $T_K$ (Field, 1959). Veil components A and B offer an unusual opportunity to study this effect since both $T_s$ and $T_K$ are independently known along the line-of-sight to the Trapezium. In particular, $T_s \approx 90$ K and 135 K for components A and B, respectively (Abel et al. 2006), while $T_K \approx 50$ K and 60 K for components A and B, respectively (Abel et al. 2016, see above). Therefore, the HI levels have been pumped in both components, and more so in component B, as one would expect if this component is closer to the Trapezium and subject to a higher stellar uv flux. However, it is not possible to be more quantitative about the pumping effect. The magnitude of the effect depends upon the unknown strengths of the Lyman lines in the stellar SED (Abel et al. 2016).

*4.3.2. Magnetic Fields in the H I Components – Precursors and/or Products of Star Formation*

Magnetic field strengths in Veil HI components A and B are anomalous compared to those measured elsewhere in the ISM in three ways: (*a*) The ratio $B/N(H)$ is 3-4 times higher in these components than it is in other regions in the ISM of comparable $N(H)$, as indicated in section 4.2 and Figure 8. Equivalently stated, $\lambda \propto N(H)/B$ is 3-4 times *lower*, hence, the gas is subcritical. (*b*) Component A is uniquely sub-Alfvenic, that is, turbulent energy is much weaker than magnetic energy, also described in section 4.2. (*c*) The ratio $B/n(H)$ in component A is about a factor of 5 higher than it is in other regions of the ISM with comparable $n(H)$. This third anomaly is based upon a statistical analysis of Zeeman effect data done by Crutcher et al. (2010). The analysis suggests $B_{tot}$ in the ISM ranges from near zero to a maximum value $B_{tot}^{max}$, where $B_{tot}^{max}$ is a function of $n(H)$. For $n(H) < n_0(H) \approx 300$ cm$^{-3}$, $B_{tot}^{max} \approx 10$ µG. For $n(H) > n_0(H)$, $B_{tot}^{max} \approx 10 \times [n(H)/n_0(H)]^{0.65}$ µG. This exponent (2/3, more specifically) is expected for isotropic collapse of a flux-frozen cloud (Mestel 1966). Observed values of $log\ B_{los}$ as a function of $log\ n(H)$ are plotted in Figure 1 of Crutcher et al. (2010). The solid line in this figure (horizontal and diagonal) represents the best fit to the Zeeman effect data for $B_{tot}^{max}$. Given the model-derived density for HI component A ($n(H) \approx 10^{2.5}$ cm$^{-3}$, section 4.3), the best-fit line predicts $B_{tot}^{max} \approx 10$ µG, one-fifth the measured $B_{los} \approx 50$ µG. The challenge presented by these three magnetic anomalies is to explain them in terms of the history of the Veil HI gas. This history begins with gravitational contraction of OMC-1 prior to star formation. The history continues with the effects of a strong stellar radiation field after star formation. Magnetic fields in the Veil today may be precursors and/or products of star formation in OMC-1. We explore these two possibilities below.

We first consider Veil HI magnetic fields as *precursors* to star formation. We focus on HI component A since it is less likely to have been affected by the Trapezium stellar environment (section 4.3.1). The higher ratio $B/n(H)$ in component A (anomaly *c* above) could be an effect of *reconnection diffusion*, assuming the Veil precursor gas existed in a low turbulence environment. Reconnection diffusion is proposed as a way to remove magnetic flux from turbulent clouds. This process may explain the insensitivity of $B$ to $n(H)$ for $n(H) < 10^{2-3}$ cm$^{-3}$ (Lazarian et al. 2012). In this scenario, field strengths remain roughly constant as cloud material gathers into a GMC since reconnection diffusion removes magnetic flux. As $n(H)$ increases, field strengths eventually begin to rise once $n(H)$ reaches a critical value $n_o(H)$. The rise in field strength begins when the reconnection diffusion timescale ($\propto 1/\Delta V_{turb}$, where $\Delta V_{turb}$ is the turbulent velocity dispersion) exceeds the free fall timescale ($\propto 1/n(H)^{\frac{1}{2}}$). Therefore, $n_o(H)$ is proportional to $\Delta V_{turb}^2$. If $\Delta V_{turb}$ was comparatively low in the OMC-1 precursor gas, then the magnetic field began rising at a comparatively low value of $n_o(H)$, leading eventually to a higher ratio $B/n(H)$, as observed in Veil component A. Since the ratio $B/n(H)$ is higher in component A by a factor $\approx 5$, we estimate that $n_o(H)$ for the precursor gas $\approx 30$ cm$^{-3}$. That is, reconnection diffusion ceased to remove magnetic flux in the Veil precursor cloud once $n(H)$ reached 30 cm$^{-3}$, and then, presumably, $B$ increased as $n(H)^{0.65}$. This behavior can be visualized by a modification to Figure 1 of Crutcher et al. (2010). In this plot of $log\ B_{los}$ vs. $log\ n(H)$, we place a horizontal line at 10 µG for $n(H) < 30$ cm$^{-3}$. For $n(H) > 30$ cm$^{-3}$, we place a diagonal line representing $B \propto n(H)^{0.65}$, the same slope actually shown in the figure for $n(H) > 300$ cm$^{-3}$. With this modification, the figure now predicts $B_{tot}^{max} \approx 50$ µG for the estimated $n(H)$ of HI component A. This field strength is (by design) the same as the Zeeman effect measurement for $B_{los}$. Since $n_0(H)$ in the Veil precursor gas was about 1/10$^{th}$ the usual value for the ISM (i.e. 30 vs. 300 cm$^{-3}$), the turbulent velocity in the Veil precursor gas was about 1/3$^{rd}$ the usual value, given that $n_0(H) \propto \Delta V_{turb}^2$, as noted above. The same reconnection diffusion process can also account for the higher ratio $B/N(H)$ in the Veil (anomaly $a$ above). For the ensemble of non-Veil Zeeman effect data, $B_{tot}^{max} \propto N(H)$ once $N(H) > N_0(H) \approx 10^{22}$ cm$^{-2}$ (Figure 8). Since the ratio $B/N(H)$ is higher for Veil component A by a factor $\approx 4$, we estimate that $N_0(H)$ for the gas that eventually became part of this component $\approx 2.5 \times 10^{21}$ cm$^{-2}$.

We conclude that reconnection diffusion in a low turbulence environment can explain the higher ratios $B/n(H)$ and $B/N(H)$ observed in the Veil HI component A compared to these ratios in the ensemble of non-Veil Zeeman data. In such a low turbulence environment, field strengths in OMC-1 begin rising at lower values of $n(H)$ and $N(H)$ than they do in more typical interstellar environments. The low turbulence hypothesis is consistent with the finding that the Alfvenic Mach number $\ll 1$ in component A, a unique property among regions probed by the Zeeman effect to date (section 4.2). The origin of low turbulence in OMC-1 could be related to the distance of the cloud above the galactic plane (150 pc). Molecular gas so high above the plane may be less subject to the input of mechanical energy from adjacent star formation and supernovae than molecular gas closer to the plane.

Now we consider Veil HI magnetic fields as *products* of star formation, that is, the effects of nearby early-type stars on the Veil magnetic fields. Pellegrini et al. (2007) describe a model in which magnetic pressure in PDRs resists the momentum of absorbed starlight (FUV and EUV) and the pressure of adjacent hot (10$^6$ K), X-ray emitting gas from shocked stellar winds. (See, also, Ferland 2009). In this model, the PDR is compressed by starlight momentum and hot gas pressure. As a result, the magnetic field is also compressed, raising the field strength and magnetic pressure. Since magnetic pressure is proportional to $B^2$, the magnetic pressure increases faster than the gas pressure for most geometries. Eventually, the magnetic pressure dominates the total pressure in the PDR and establishes an approximate magnetostatic equilibrium. Note that this process requires some component of $B$ to lie in the plane of the Veil PDR (i.e. perpendicular to the line of sight) since it is this field component that can be amplified by compression of the PDR. The Pellegrini et al. model also results in an increase in $B/N(H)$, as observed in Veil components A and B, since compression of the PDR increases $B$ but not $N(H)$. Pellegrini et al. used the spectral synthesis code Cloudy (Ferland et al. 1998; Ferland et al. 2013) to model the M17 PDR along the lines described above. They successfully reproduced magnetic field strengths previously observed with the VLA via the HI Zeeman effect. Note that this model assumes some component of $B$ parallel to the illuminated face of the PDR.

To apply the Pellegrini et al. concepts to the Veil, we use an approximate technique (as opposed to a full Cloudy model) to estimate field strengths in the Veil HI regions. We estimate field strengths by equating magnetic pressure ($B^2/8\pi$) to radiation pressure plus the pressure of X-ray emitting gas. Radiation pressure is given by $F/c$, where $F$ is the flux (erg cm$^{-2}$ s$^{-1}$) integrated over a specified wavelength band and incident upon the PDR. We distinguish between $F_{EUV}$ (flux of H-ionizing radiation) and $F_{FUV}$ (uv flux of non H-ionizing radiation). Also, the pressure of the hot gas $P_x = nkT_x$, there $T_x$ is the kinetic temperature. Equating magnetic pressure to radiation pressure plus hot gas pressure, we arrive at the predicted PDR field strength $B_{PDR} = (8\pi)^{\frac{1}{2}} (F_{EUV}/c + F_{FUV}/c + P_x)^{\frac{1}{2}}$, all in cgs units This relation is equivalent to equation 7 of Pellegrini et al. except we also include the effects of the non-ionizing radiation $F_{FUV}$. To estimate $F_{EUV}$ and $F_{FUV}$ for the Orion Veil, we use the Trapezium SED derived by Abel et al. (2016) and assume the Veil is 2 pc from the Trapezium (section 4.3.1). Thus, we find $F_{EUV} \approx 0.54$ erg cm$^{-2}$ s$^{-1}$, and $F_{FUV} \approx 1.5$ erg cm$^{-2}$ s$^{-1}$. We take $P_x/k = 10^6$ K cm$^{-3}$, based on the Orion X-ray observations of Gudel et al. 2008. This estimate for $P_x$ is rather uncertain because the properties of the hot gas immediately adjacent to the Veil are not well constrained by the X-ray observations. With these parameters, we predict $B_{PDR} \approx 70$ μG, including radiation and hot gas pressure, and $B_{PDR} \approx 40$ μG if only radiation pressure is included. These magnetic field strengths are comparable to those measured in HI components A and B. We conclude that a model in which the Veil PDR is magnetically supported against stellar radiation and hot gas pressures is consistent with the HI Zeeman effect observations. Mayo & Troland (2012) applied this same concept to the PDR associated with H$^+$ region DR22.

The implications of such a magnetically supported PDR model, along the lines described by Pellegrini et al. (2007), are two fold. First, the model suggests that the presently observed magnetic fields in the HI Veil may have been increased by the presence of nearby early type stars. If the initial field strengths in the Veil PDR were insufficient to support it against radiation pressure and X-ray emitting gas from the Trapezium, then the PDR would have been compressed, raising the field strength until equilibrium was achieved. This process would increase the ratio $B/N(H)$, as observed; however, it would not increase the ratio $B/n(H)$. Of course, OMC-1 cloud evolution may have led to equilibrium level field strengths *prior to* star formation as described above. In such a case, no further compression of the PDR and enhancement of $B$ would have been necessary. The other more general implication of the magnetically supported PDR model is that magnetic fields can play an important role in establishing physical conditions within these regions. Since magnetic pressure is dominant in such a PDR model, gas pressure, hence, gas density, is lower than it would be in a non-magnetic environment. As a result, the magnetized H$^0$ (for a given N(H)) extends more deeply into the PDR. That is, the dominance of magnetic pressure in the H$^0$ gas significantly affects physical conditions there, creating a deeper, lower density atomic layer than expected in the non-magnetic case. If so, then PDR models that do not account for magnetic pressure may significantly underestimate the depths and overestimate the gas densities in the H$^0$ regions.

### 4.4. The Location of Orion-S

Orion-S is one of two high-mass molecular cores in OMC-1; Orion-KL is the other. Orion-S is 90" south of Orion-KL and about 60" southwest of the Trapezium cluster. (See section 3.3 and Figure 2.) Like Orion-KL, Orion-S is clearly visible in images of submillimeter dust emission (Lis et al. 1998; Johnstone & Bally 1999) and in molecular emission line images (e.g. Peng et al. 2012). The 850 μm dust emission image of Johnstone & Bally (see contour image in Peng et al. 2012) reveals an Orion-S core approximately 40" (0.09 pc) in diameter, with $N(H) \approx 2$-$4 \times 10^{24}$ cm$^{-2}$ (Av $\approx$ 800-1600 mag). The factor of two uncertainty reflects uncertainties in theoretically calculated values of the dust opacity $\kappa_\nu$. Other uncertainties exist in this $N(H)$ estimate, for example, in $T_{dust}$ which we have taken as 40 K (O'Dell et al. 2009). The implied total mass for Orion-S is 130 – 260 solar masses, the average density, assuming spherical symmetry, is 1-2 × 10$^7$ cm$^{-3}$. The virial mass of Orion-S, for an observed $\Delta V \approx 5$ km s$^{-1}$ and radius of 0.05 pc, is about 150 solar masses, suggesting that the core is close to virial stability.

It is logical to imagine that Orion-S lies behind the main Orion $H^+$ region, just like the Orion-KL core. For example, there is no obvious optical extinction feature associated with this optically opaque core, just as there is none for Orion-KL. However, the situation for Orion-S is more complicated owing to several discoveries reported in the previous literature. Johnston et al. (1983) and Mangum et al. (1993) found 6 cm (4.8 GHz) $H_2CO$ absorption toward Orion-S. The present OH data also reveal absorption (section 3.3). The present data also show H I absorption toward Orion-S, and the data of vdWGO13 show this same absorption at higher spatial resolution (identified by these authors as H I component H). We illustrate the clear kinematical association between OH and H I absorption and Orion-S molecular emission in Figure 10. This figure presents H I and OH optical depth profiles from the present work and a CS (2-1) emission profile from Tatematsu et al. (1998). See the figure caption for beamwidths and channel separations. All three profiles are toward the center position of Orion-S. The OH and CS profiles have nearly identical center velocities (7 km s$^{-1}$) and velocity widths. The H I profile has a weak component on the high velocity side of the main velocity component that appears to match the OH and CS features. (The low velocity side of this weak H I component, at velocities 3 - 7 km s$^{-1}$, is presumably obscured by the main H I velocity component.) Evidently, a source of radio continuum emission lies behind all or part of Orion-S in order to account for the HI, OH and $H_2CO$ absorption. Most likely, this radio continuum emission comes from an $H^+$ region. Another indication of the complexity of the Orion-S core comes from anomalies in the optical extinction reported by O'Dell et al. (2009). These anomalies suggest that part of the radio continuum observed toward Orion-S comes from an ionized layer from which we receive no H$\alpha$ light. Presumably, this $H^+$ layer lies behind an optically opaque layer associated with Orion-S. In short, previous and present observations point to the presence of two distinct layers of $H^+$ along the line of sight to Orion-S. These layers are separated by an optically opaque region that consists of all or part of Orion-S.

Apart from kinematics (Figure 10), there is an intriguing spatial correspondence between the H I absorption and the Orion-S core defined by dust emission. The vdWGO13 data offer a high spatial resolution image of this absorption. In Figure 11 we overlay contours of the 850 μm dust emission (dashed contours, Johnstone & Bally 1999) upon an image of H I optical depth from the data of vdWGO13 (colors and light contours). The H I optical depth image has been integrated over the velocity range 8.4 to 9.7 km s$^{-1}$, a range over which optical depth associated with Orion-S is uncontaminated by unrelated velocity features (see Figure 10, top panel). The HI image has a synthesized beamwidth of 6.4". Figure 11 shows that the H I absorption arises from a partial shell just to the west of the peak in 850 μm dust emission associated with Orion-S. Although the morphologies of 850 μm dust emission and H I optical depth are far from identical, the general correspondence between the two provides strong evidence, along with the kinematical evidence (Figure 10), that the H I absorption is closely associated with Orion-S. The same conclusion holds for the $H_2CO$ and OH absorption.

O'Dell et al. (2009) propose that Orion-S is an isolated molecular core lying *in front of* the main ionization front (MIF) of the Orion Nebula. (See their Figure 5.) Therefore, the main $H^+$ region provides radio continuum absorbed by $H_2CO$, OH and HI. In this model, light from the Trapezium cluster ionizes the front face of Orion-S. As seen from the Earth, light from this ionized Orion-S layer fills in light from the main $H^+$ region that is blocked by Orion-S. Therefore, Orion-S does not create an obvious optical extinction feature even though it lies in front of the MIF. O'Dell et al. also provide a detailed spectral synthesis calculation of Orion-S, predicting the run of density, dust and gas temperatures and chemical composition into the core. This model establishes, for example, that $T_{dust}$ in the core remains above 30 K needed to keep $H_2CO$ from freezing out onto grains.

Several objections can be raised to the O'Dell et al. (2009) model, none of which excludes it. These objections are related to the need for ionized gas behind Orion-S to account for the atomic and molecular absorption. In the O'Dell et al. (2009) model, the position of Orion-S must be carefully chosen so that it does not shadow the main Orion $H^+$ region behind it. If Orion-S were to shadow this layer, then the layer would not be ionized, so it could not emit the radio continuum. Also, such a shadow would likely be visible in optical images of the Orion Nebula to the southwest of Orion-S (i.e. opposite to the direction toward the Trapezium). No such shadow is apparent (Figure 1a, southwest of position OH4 that is coincident with Orion-S). Indeed, the existence of an *isolated* molecular core in Orion, surrounded by $H^+$, seems unlikely. Most of the ionizing radiation in Orion comes from a single star, Theta $^1$C Ori. Therefore, Orion-S must cast a shadow, creating an elephant trunk of neutral gas similar to those observed in M16 (Hester et al. 1996). The postulated Orion-S elephant trunk would be seen from a different angle than those of M16. The Orion-S structure would point more nearly toward the observer in Orion and be seen against the bright background of the main $H^+$ region. This geometry would be quite unlike that of the M16 elephant trunks that lie approximately in the plane of the sky and are viewed against a darker background. As a result, the Orion-S elephant trunk might be much less visible in optical images than those of M16. Nonetheless, an Orion-S elephant trunk would have to be carefully positioned to account for $H^+$ behind the Orion-S core. With suitable choice of geometry, this background $H^+$ gas could be part of the main Orion $H^+$ region or it could lie along the skin of the elephant trunk on the side of the structure facing away from Earth. Either way, the geometry of the Orion elephant trunk would have to be rather optimally configured in order to place a significant amount of ionized gas directly behind Orion-S, while Orion-S itself lay in front of the MIF.

We now offer another explanation for the presence of HI, OH and $H_2CO$ absorption toward Orion-S that avoids the need to carefully position the molecular core in front of the MIF. In this model, the Orion-S molecular core lies entirely *behind* the MIF, just like Orion-KL. On the backside of Orion-S (i.e. facing away from Earth), an early-type star has created a small $H^+$ region whose radio continuum emission is responsible for the observed atomic and molecular absorption. The small $H^+$ region could be a blister-type region, similar to, but smaller than the Orion Nebula or it could be embedded within Orion-S. We have made a simple estimate of the type of star that could form such a small $H^+$ region. We use the formalism described by Jackson & Kraemer (1999) and references therein. Consider an $H^+$ region as large as Orion-S (0.1 pc) in the plane of the sky. Suppose that this $H^+$ region has *half* the brightness (in Jy beam$^{-1}$) as the main Orion $H^+$ region. (Were the small $H^+$ region much brighter, it would create an obvious peak in total brightness toward Orion-S, contrary to observation.) Given the assumed brightness and size of the small $H^+$ region, its total 21 cm flux density is about 4 Jy. Assuming an electron temperature $T_e \approx 8000$ K, the small $H^+$ region could be created by a B0 ZAMS star, a star far less luminous than Theta $^1$C Ori. Such a star, on the back side of Orion-S or embedded within it, would be invisible behind Av $\approx$ 1000 from Orion-S. The postulated Orion-S $H^+$ region is shown in Figure 9 as a red circle immediately behind Orion-S.

This second Orion-S model is *ad hoc* but plausible on several grounds. For one, Orion-S is well established as a star-forming region (e.g. Henney et al. 2007). Therefore, the presence of a B0 star in its vicinity would be no surprise. Also, the roughly circular region of enhanced optical extinction in the image of O'Dell & Yusef-Zadeh (2000) is easily explained as an artifact arising from background radio continuum emission of a small Orion-S $H^+$ region from which no H$\alpha$ emission is observed owing to the large optical extinction of Orion-S in front. (This is the anomalous extinction phenomenon cited above and described by O'Dell et al. 2009.) Finally, the partial shell of H I optical depth adjacent to Orion-S (see above) can be naturally explained as arising in a partial shell of $H^0$ gas lying just outside the small Orion-S $H^+$ region. In this simple model, the Orion-S core is just like the Orion-KL core, behind the main Orion $H^+$ region. If this model is correct, then the Orion-S core may lie just barely behind the main Orion $H^+$ region. Orion-S coincides spatially with a local maximum or bump in the H+ region surface that points toward Earth. This bump appears in the 3-D map of Wen & O'Dell (1995). These authors note that the bump could be caused by dense gas behind the $H^+$ region that retards the advance of the ionization front. This dense gas, of course, would be the Orion-S molecular core.

In short, two different models of Orion-S can account for existing observations, especially the presence of atomic and molecular absorption associated with the core. One model (O'Dell et al. 2009) places Orion-S in *front* of the MIF, as an isolated neutral region immersed in $H^+$ gas. It is only a slight modification of this model (described above) to replace the isolated core with an elephant trunk having the core at its tip. The other model (present work) places Orion-S *behind* the MIF (like Orion-KL) with yet another, smaller $H^+$ region behind Orion-S. Both models require *ad hoc* assumptions. However, we favor the placement of Orion-S behind the MIF since this placement avoids the necessity of rather specific geometry of an isolated molecular region or elephant trunk. Future high spatial resolution images of the atomic and molecular absorption regions associated with Orion-S might assist in distinguishing between these two models. For example, if the morphologies of H I and molecular absorption are found to trace a partial shell around the molecular core (a circumstance hinted at in Figure 11), this result would strengthen the hypothesis that a small $H^+$ region lies behind Orion-S, and a partial shell of neutral gas surrounds the ionized gas. That is, this result would strengthen the model placing Orion-S behind the MIF.

## 5. SUMMARY AND CONCLUSIONS

The Orion Veil is a mostly atomic PDR that lies about 2 pc in front of the Orion Trapezium stars (as viewed from Earth) and nearly perpendicular to the line of sight. The Veil covers all of the Orion Nebula and extends at least as far northeast as the smaller $H^+$ region M43. Radio, optical and UV absorption lines from the Veil provide an unusual opportunity to study PDR physical conditions. These lines reveal two principle H I velocity components, labeled A and B. We report aperture synthesis observations of the Zeeman effect in 21 cm H I absorption lines and in 18 cm (1665 and 1667 MHz) OH absorption lines across the Veil. The H I data yield independent images of the line-of-sight magnetic field strength $B_{los}$ in components A and B with a synthesized beamwidth of 25". The OH data provide a measurement of $B_{los}$ in an optically obscured molecular region of the Veil along the northeast edge of the Orion Nebula. The synthesized beamwidth of the OH observations is 40".

From the H I Zeeman data, we find $B_{los} \approx$ -50 to -75 µG across much of the Veil, rising to about -300 µG toward the Northeast Dark Lane, a region of high optical obscuration along the northeast edge of the Orion Nebula (section 3.4). Images of $B_{los}$ for H I velocity components A and B (Figures 5 and 6) are very similar. This similarity suggests a close association between the atomic layers responsible for the two components. From the OH Zeeman data, we find $B_{los}$ = -350 ± 47 µG in molecular gas of the Northeast Dark Lane (section 3.5) at a position where $B_{los,HI}$ is -183 ± 32 µG in H I component A.

We assess the importance of thermal, turbulent, magnetic and gravitational energies in the Veil and compare these energies with energies in other locales of the ISM (section 4.2, Tables 4a and 4b). H I components A and B have mild supersonic turbulence ($M_{turb} \approx 3$), similar to the CNM. However, molecular gas in the Veil has much stronger supersonic turbulence ($M_{turb} \approx 4$-6), similar to molecular gas in other regions of high mass star formation. Thermal energy is insignificant in H I components A and B compared to magnetic energy ($\beta_{therm} \ll 1$), as it is in most of the ISM. An approximate balance exists between turbulent and magnetic energies in much of the atomic and molecular gas of the Veil, just as it does in other regions the ISM probed with the Zeeman effect (trans-Alfvenic turbulence, $\beta_{turb} \approx 1$, $M_A \approx 1$). However, H I component A is uniquely sub-Alfvenic ($\beta_{turb} \approx 0.02$, $M_A \approx 0.1$); magnetic energy strongly dominates turbulent energy. We estimate the mass-to-flux ratio (ratio of gravitational to magnetic energies) of gas in the Veil. Veil H I components A and B are magnetically *subcritical* ($\lambda < 1$, magnetically dominated), similar to the CNM where $n(H)$ is about 1 dex lower. The molecular gas of the Veil, where sampled by the Zeeman effect, appears magnetically *critical* to *supercritical* ($\lambda \geq 1$, gravitational energy comparable to or greater than magnetic energy), like molecular gas in other regions of high and low mass star formation.

We consider the physical nature of H I components A and B (section 4.3). Like previous authors, we argue that component B is closer to the Trapezium stars than component A. (See Figure 9.) This placement explains why component B is warmer, denser and more turbulent than component A. Component B may consist of post shock gas associated with a weak-D ionization front, its blueshift relative to component A arising from the passing shock that precedes the ionization front. This model requires the magnetic field in both components to lie nearly along the line-of-sight (i.e. perpendicular to the shock front).

Magnetic fields in H I components A and B are anomalous in the sense that they are 3 – 5 times stronger than fields in other regions of the ISM with comparable $N(H)$ and $n(H)$. See section 4.3.2 and Figure 8. One explanation for these anomalies lies in the history of OMC-1 *before* star formation. If OMC-1 was formed in an unusually low-turbulence environment (perhaps because of its position 150 pc above the galactic plane), then the process of reconnection diffusion was less efficient at removing magnetic flux from the developing cloud. As a result, gravitational contraction began at lower values of $n(H)$ and $N(H)$, leading to higher ratios $B/N(H)$ and $B/n(H)$ today. This hypothesis is consistent with the low Alfvenic Mach number in HI component A (i.e. low turbulent energy) since this gas likely samples material in OMC-1 that has yet to be disturbed by the Trapezium stars. Another explanation for the magnetic anomalies of H I components A and B lies in events that occurred *after* star formation. Pressure from starlight and from adjacent hot ($10^6$ K) gas can compress the PDR, raising field strengths within it until a magnetostatic equilibrium is achieved. We find that field strengths in components A and B are consistent with such an equilibrium.

Finally, we consider the location of the Orion-S molecular core (section 4.4). This massive core ($M \approx 200$ $M_{sun}$, $N(H) \approx 3 \times 10^{24}$ cm$^{-2}$) is similar in nature to Orion-KL. However, the Orion-S location is unclear. The existence of HI and molecular absorption lines spatially and kinematically associated with Orion-S (Figures 10, 11) establish that a source of radio continuum lies behind the core. O'Dell et al. (2009) propose that Orion-S is an isolated core lying in front of the main ionization front (MIF) of the $H^+$ region. We note several possible objections to this idea, none of which rules it out. We prefer another location for Orion-S, placing the core *behind* the MIF, with another, smaller $H^+$ region behind or embedded within Orion-S. Future high spatial resolution observations of the atomic and molecular absorption lines may help distinguish between these two models.

We gratefully acknowledge the careful reading of an early version of this manuscript by C. R. O'Dell as well as his numerous comments and suggestions. This work has been supported in part by NSF grant AST 0908841 to THT. We also acknowledge the referee's thorough reading of the manuscript and his or her constructive suggestions.

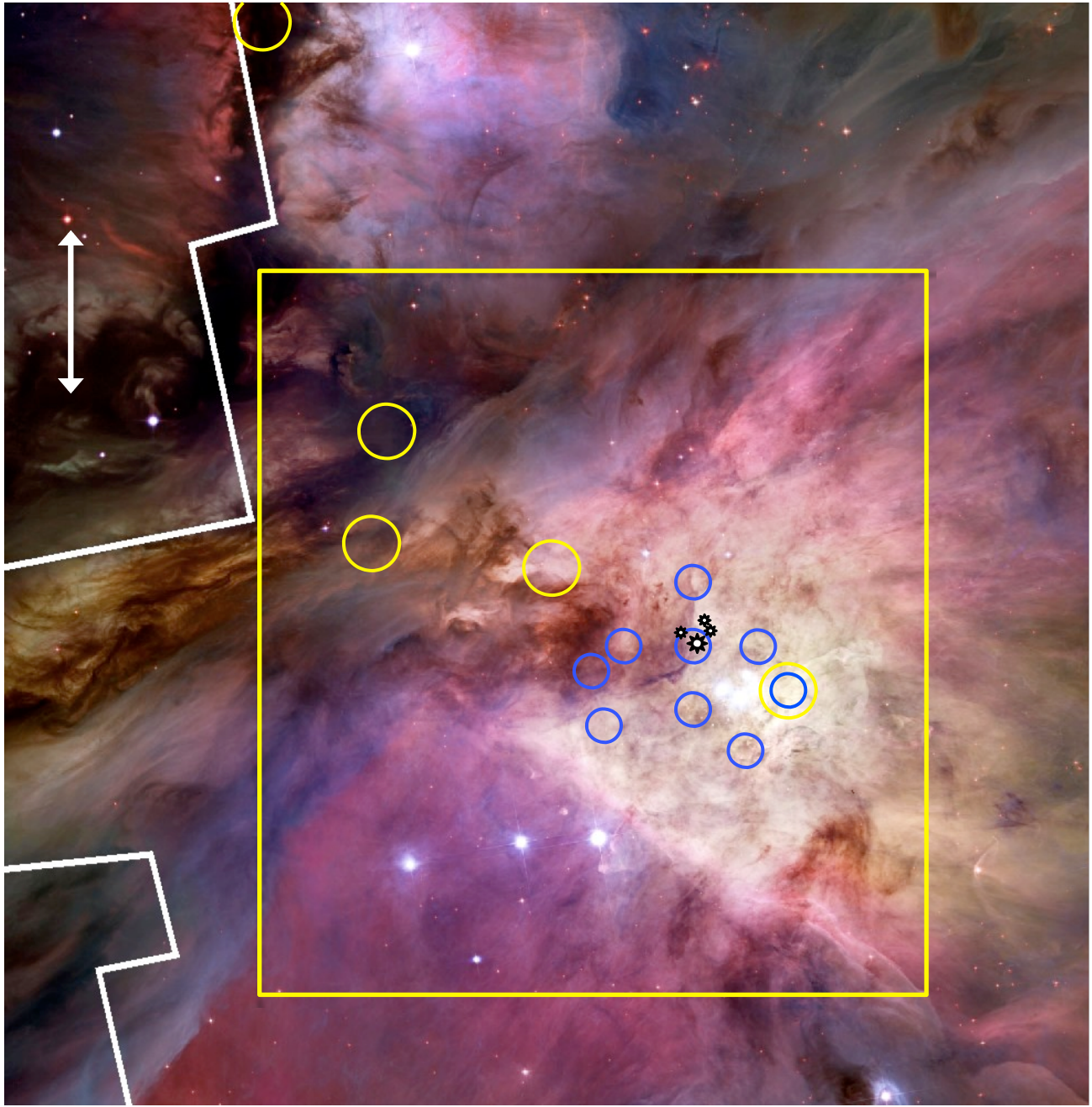

**Figure 1a.** Optical image of $H^+$ regions M42 (center) and M43 (near top) taken from Henney et al. (2007). The Trapezium stars are denoted by black symbols, the largest symbol represents Theta 1C Ori. The vertical arrow is 2' (0.25 pc) long. Blue circles denote representative positions HI1 – HI9 where unsaturated H I data were extracted (Table 3); the circles are 25" in diameter, corresponding to the 21 cm synthesized beamwidth. Yellow circles denote representative positions OH1 – OH5 where OH data were extracted (Table 2, Figures 2 and 3); the circles are 40" in diameter, corresponding to the 1665 and 1667 MHz synthesized beamwidth. The yellow rectangle indicates the field of view in Figures 5 and 6.

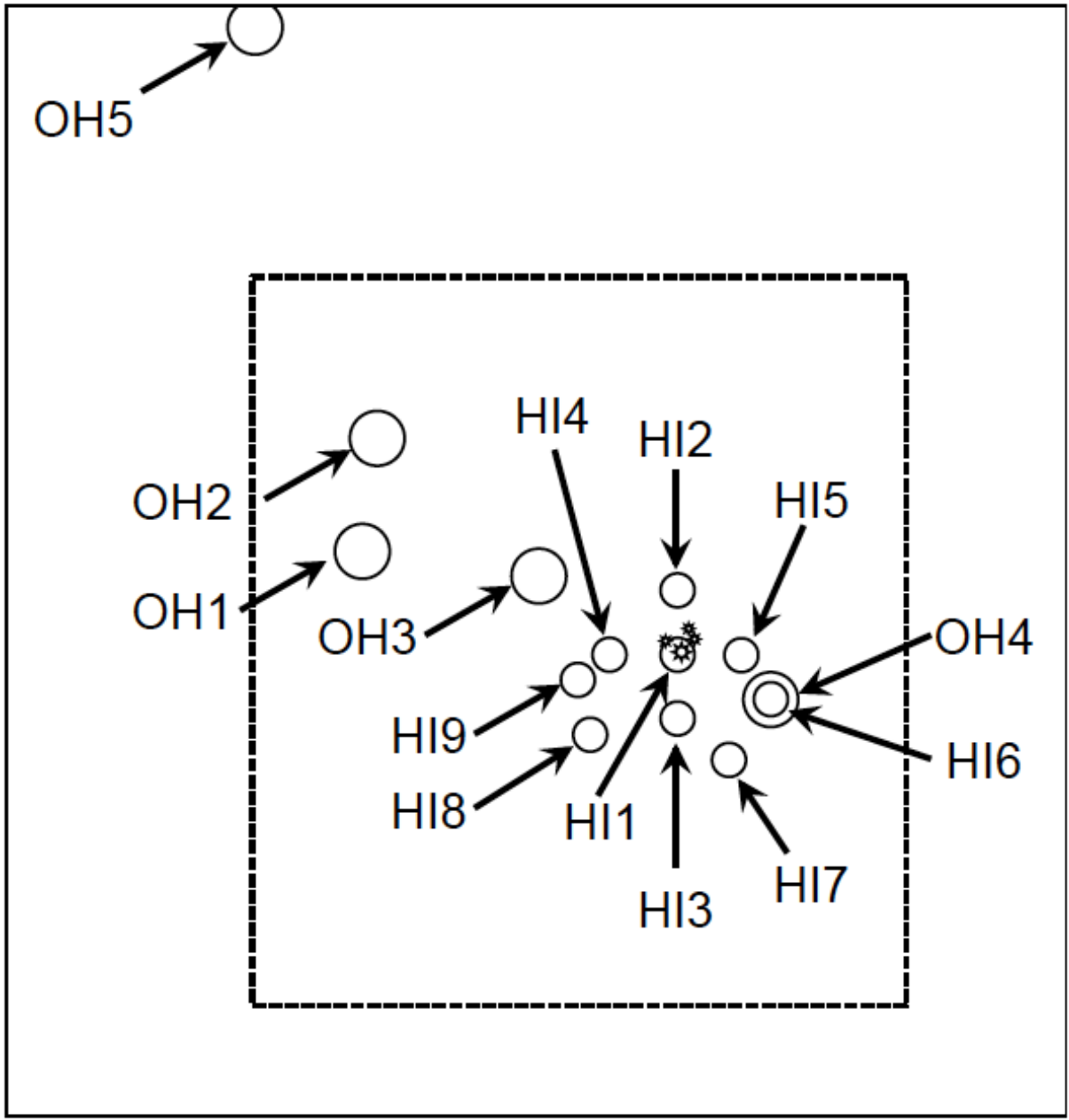

**Figure 1b.** Finder chart for positions shown in Figures 1a, 2, 5 and 6 and listed in Tables 2 and 3. Outer solid rectangle indicates full field of view of Figure 1a. Inner dashed rectangle indicates the field of view in Figures 5 and 6; it is the same as the yellow rectangle of Figure 1a.

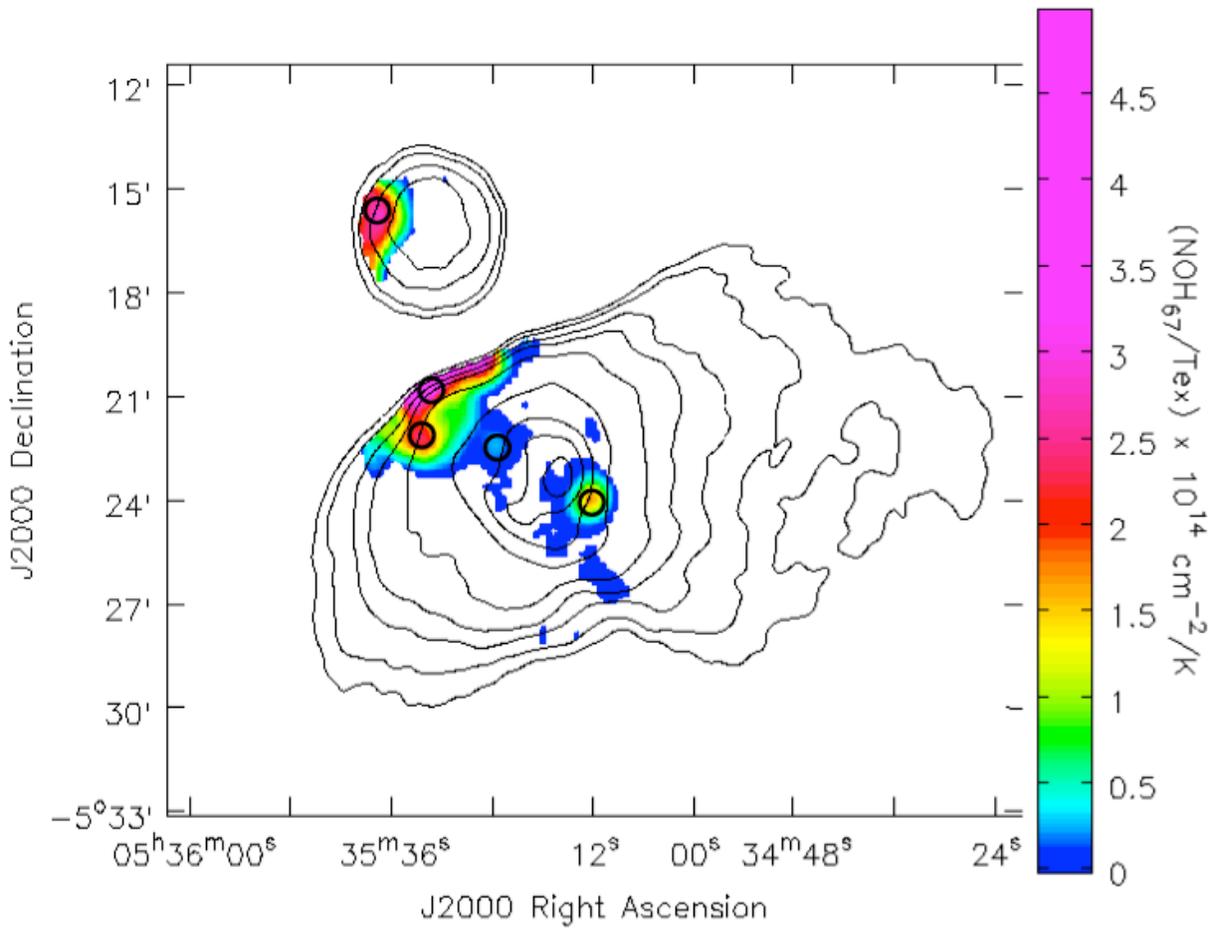

**Figure 2.** (Colors) Integrated 1667 MHz OH optical depth in units of $[N(OH)/T_{ex}] \times 10^{14}$ cm$^{-2}$ K$^{-1}$, where $T_{ex}$ is the excitation temperature. Black contours represent the 1667 MHz continuum brightness; isolated contours in the northeast sector of the image are for M43. Synthesized beam size is 40". Black circles are 40" in diameter; they denote positions OH1 – OH5 for which OH 1667 MHz optical depth profiles are shown in Figure 3 and positions and other data given in Table 2. Also, see finder chart in Figure 1b.

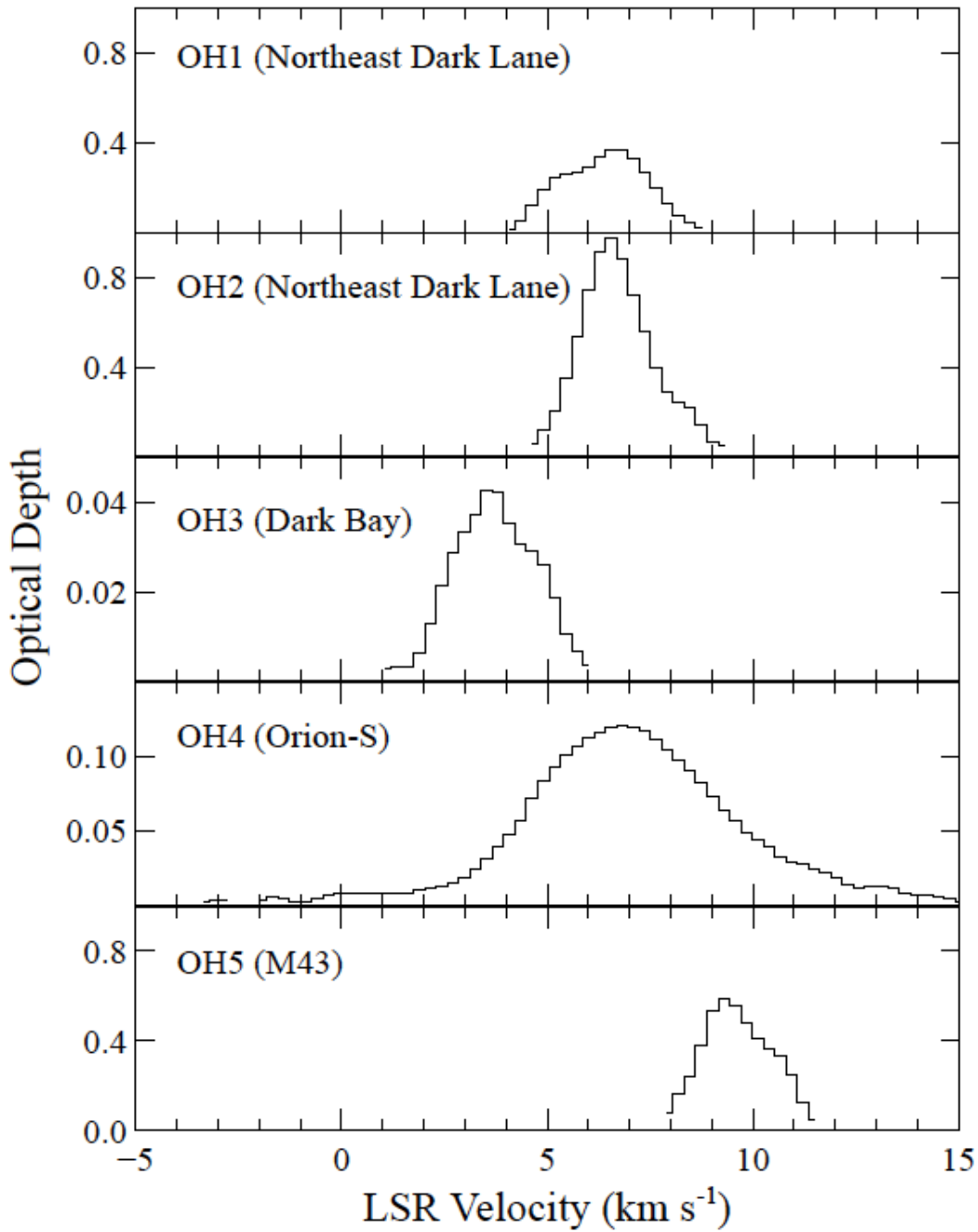

**Figure 3.** 1667 MHz OH optical depth profiles for positions OH1 – OH5 listed in Table 2 and shown as yellow circles in Figure 1a, larger black circles in Figure 1b, and black circles in Figure 2. Synthesized beamwidth is 40"; channel separation is 0.27 km s$^{-1}$.

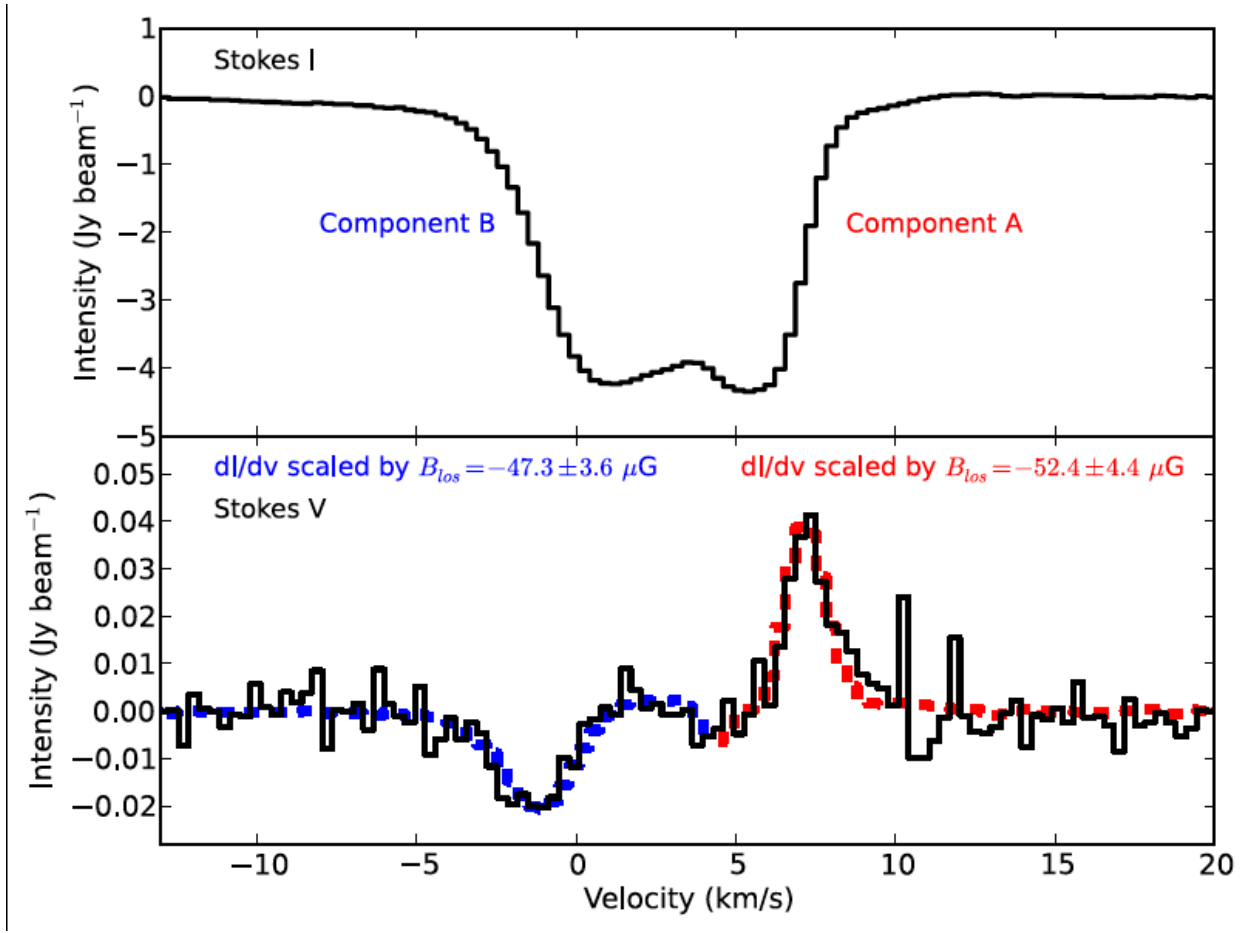

**Figure 4.** Stokes *I*/2 and Stokes *V*/2 profiles toward the Trapezium stars (position 1 in Table 3) for 21 cm H I line. The dashed profiles overlaid upon the Stokes *V*/2 profile are derivatives of the Stokes *I*/2 profiles, scaled separately for the best fit values of $B_{los}$ in H I components A and B, -52.4 ± 4.4 µG and -47.3 ± 3.6 µG, respectively. Velocities are LSR. Synthesized beamwidth is 25"; channel separation is 0.32 km s$^{-1}$.

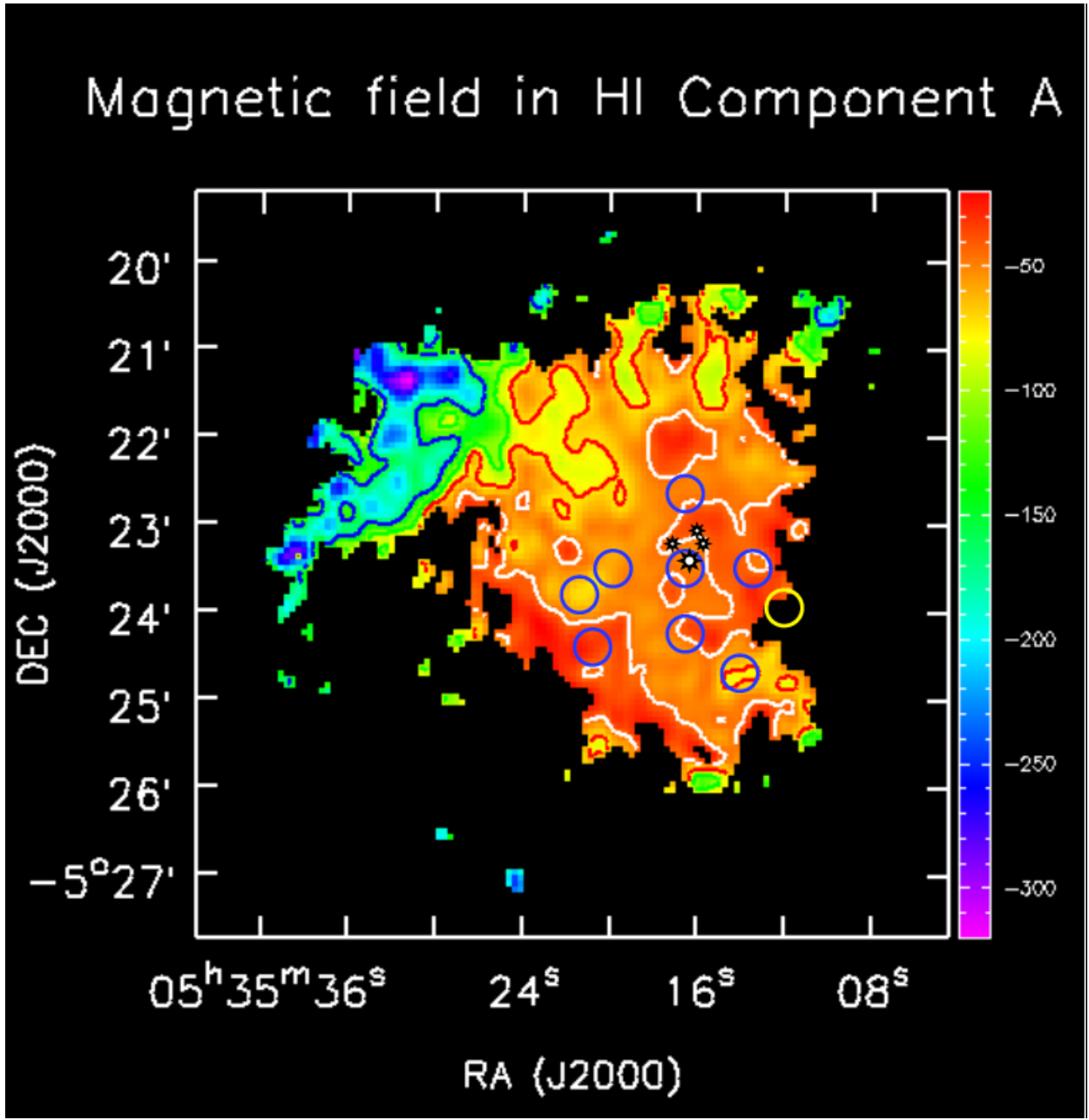

**Figure 5.** Image of $B_{los}$ (colors and contours) for H I component A. Contours are at $B_{los}$ = -45 µG (white), -75 µG (red), -100 µG (green) & -150 µG (blue). The image has been masked to include only those pixels for which $|B_{los}|/\sigma(B_{los}) > 3$. The Trapezium stars are denoted by black symbols as in Figure 1a. Circles of 25" diameter (synthesized beamwidth) denote positions HI1 – HI9 in Table 3 for which values of $B_{los}$ are given. For comparison purposes, the field of view of this image is shown in Figures 1a and 1b as dashed rectangles. Also, see finder chart in Figure 1b.

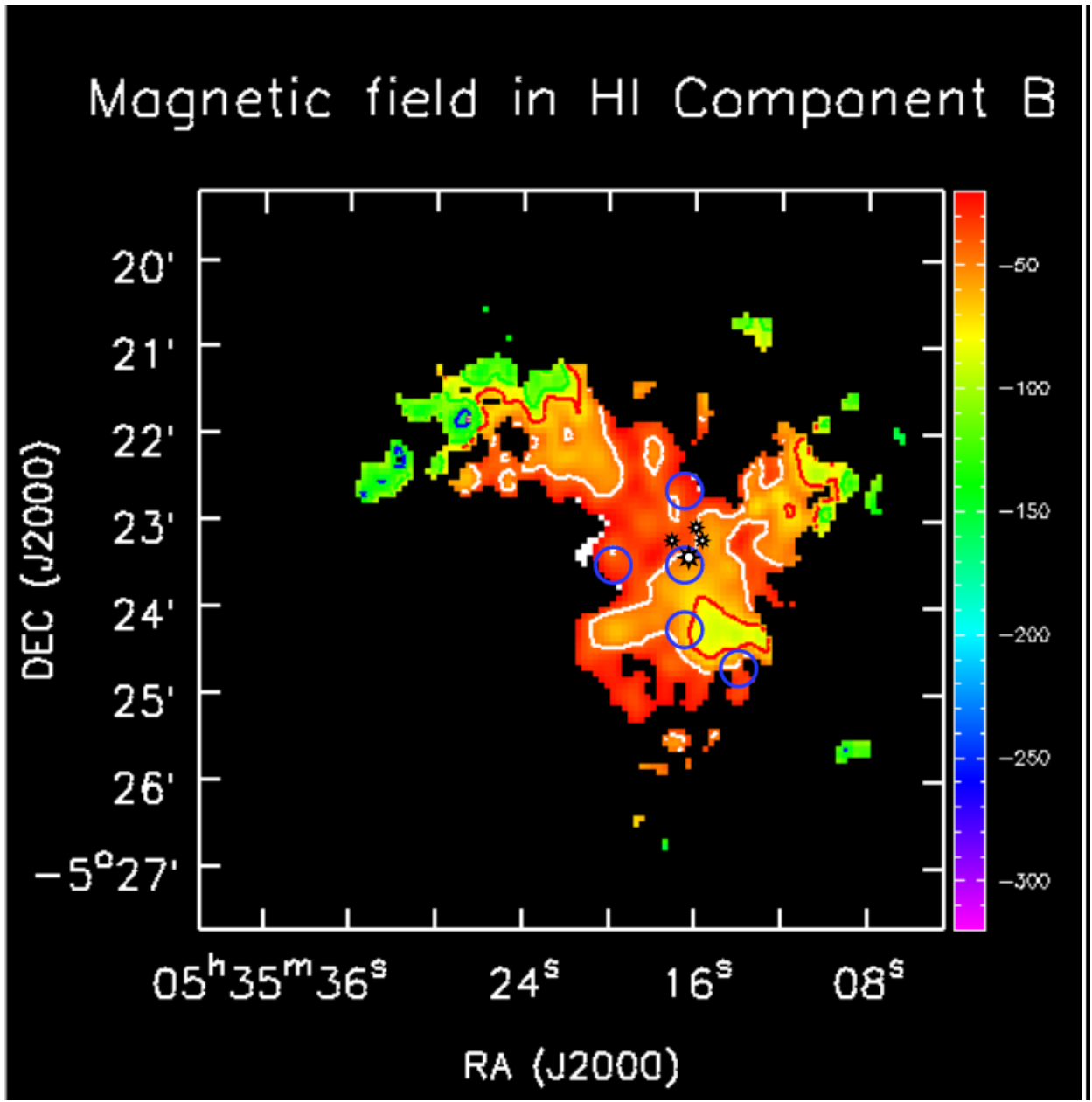

**Figure 6.** Image of $B_{los}$ (colors and contours) for H I component B. Same as Figure 5 except circles denote positions HI1 - HI4 and HI7 in Table 3 for which values of $B_{los}$ are given.

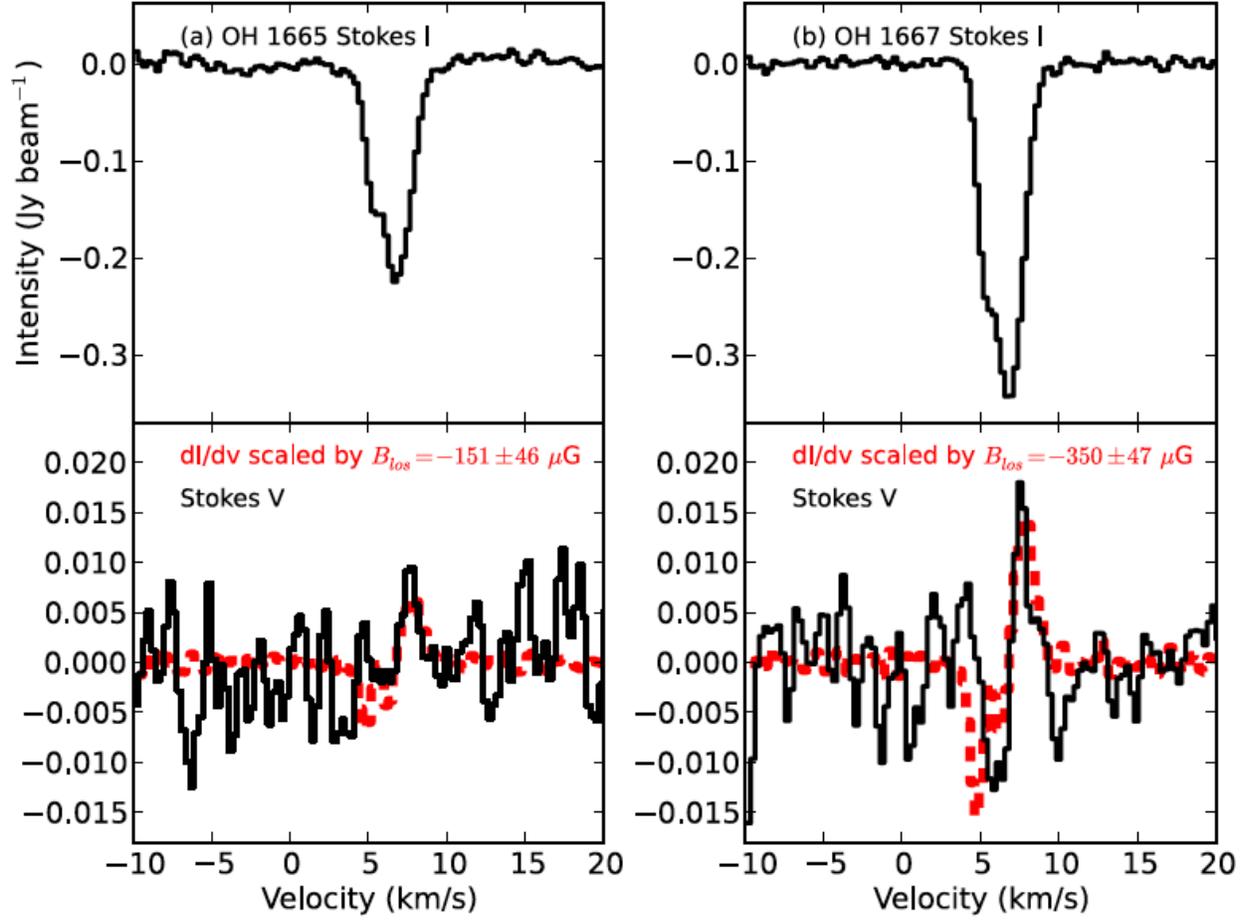

**Figure 7.** Stokes I/2 and Stokes V/2 profiles for the 1665 MHz OH line (left) and for the 1667 MHz line (right) at position OH1. The red dashed lines overlaid upon the Stokes V/2 profiles are derivatives of the Stokes I/2 profiles, scaled separately for the best-fit values of $B_{los}$ in the two OH lines. Velocities are LSR. Synthesized beamwidth is 40"; channel separation is 0.27 km s$^{-1}$.

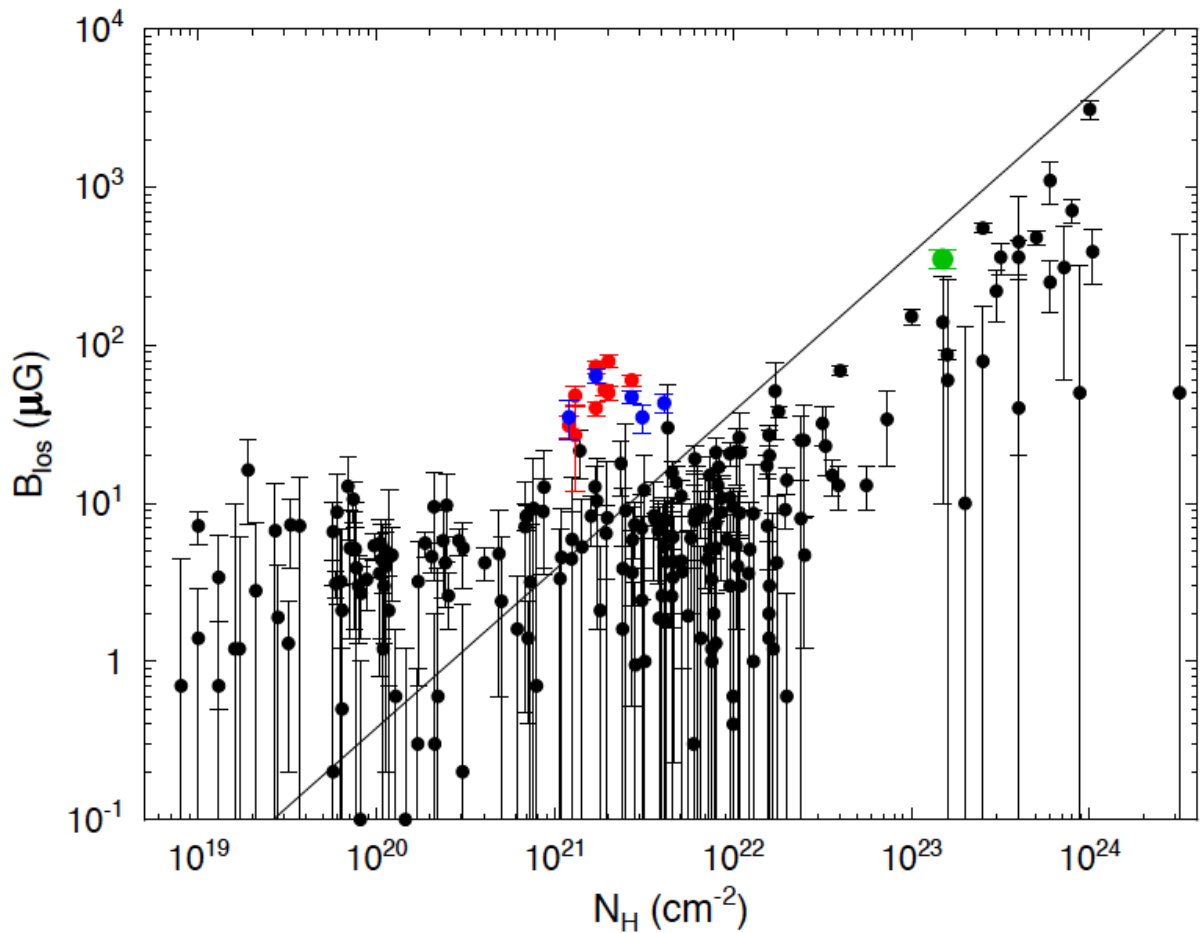

**Figure 8.** Line-of-sight magnetic fields $B_{los}$ from Zeeman effect measurements plotted against H column densities $N(H) = N(H^0) + 2N(H_2)$. Original plot taken from Crutcher (2012) who lists references. Blue dots indicate $B_{los}$ in H I component A, red dots indicate $B_{los}$ in H I component B, and the green dot indicates $B_{los}$ in the OH-absorbing molecular gas of the Northeast Dark Lane.

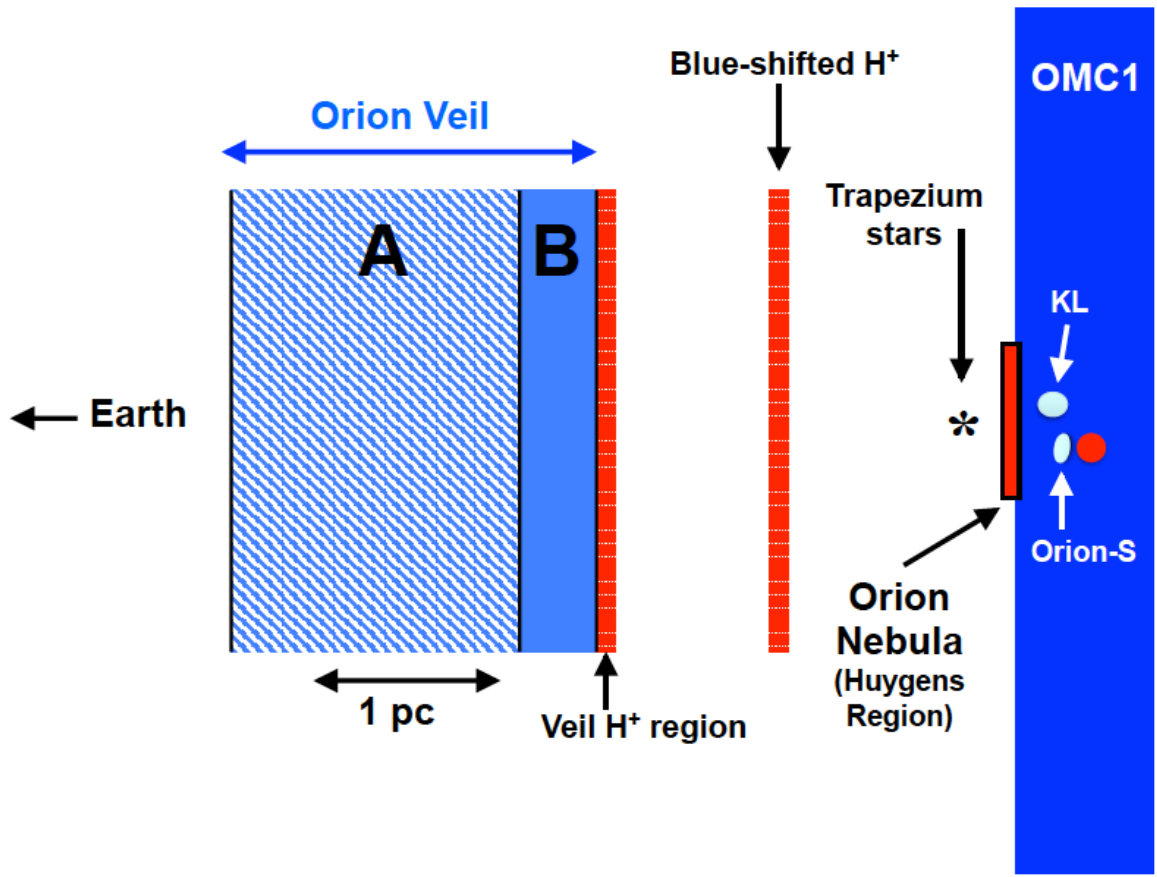

**Figure 9.** Cartoon of the Orion Nebula environment seen *sideways* (i.e. edge-on) with the Earth at left and north up. Blue regions are neutral gas, with the darkest blue representing molecular gas of OMC-1. Red regions are $H^+$. Included in the diagram is the filamentary molecular cloud OMC-1 with the Orion-KL core and the Orion-S core within, the Orion Nebula $H^+$ region (bright inner Huygens region), the Trapezium stars, and the Orion Veil. The Veil is shown with its two principal $H^0$ velocity components A and B, and the presumed Veil $H^+$ region facing the Trapezium stars. Horizontal dimensions are to approximate scale (except the width of the Veil $H^+$ region for which no width estimate exists). Width estimates for components A and B are from Abel et al. (2016). The vertical dimension of the Orion Nebula Huygens Region is to scale, the OMC-1 cloud is a filament extending north and south of the diagram. The vertical extent of the Veil is arbitrary, although it must extend at least 1 pc above the Trapezium to cover M43. Possible location of the Orion-S core within OMC-1, hence, behind the Orion Nebula, is discussed in section 4.4. An $H^+$ region (filled red circle) behind Orion-S is also discussed in section 4.4. See O'Dell et al. (2009, Figure 5) for an alternate model of the location of Orion-S.

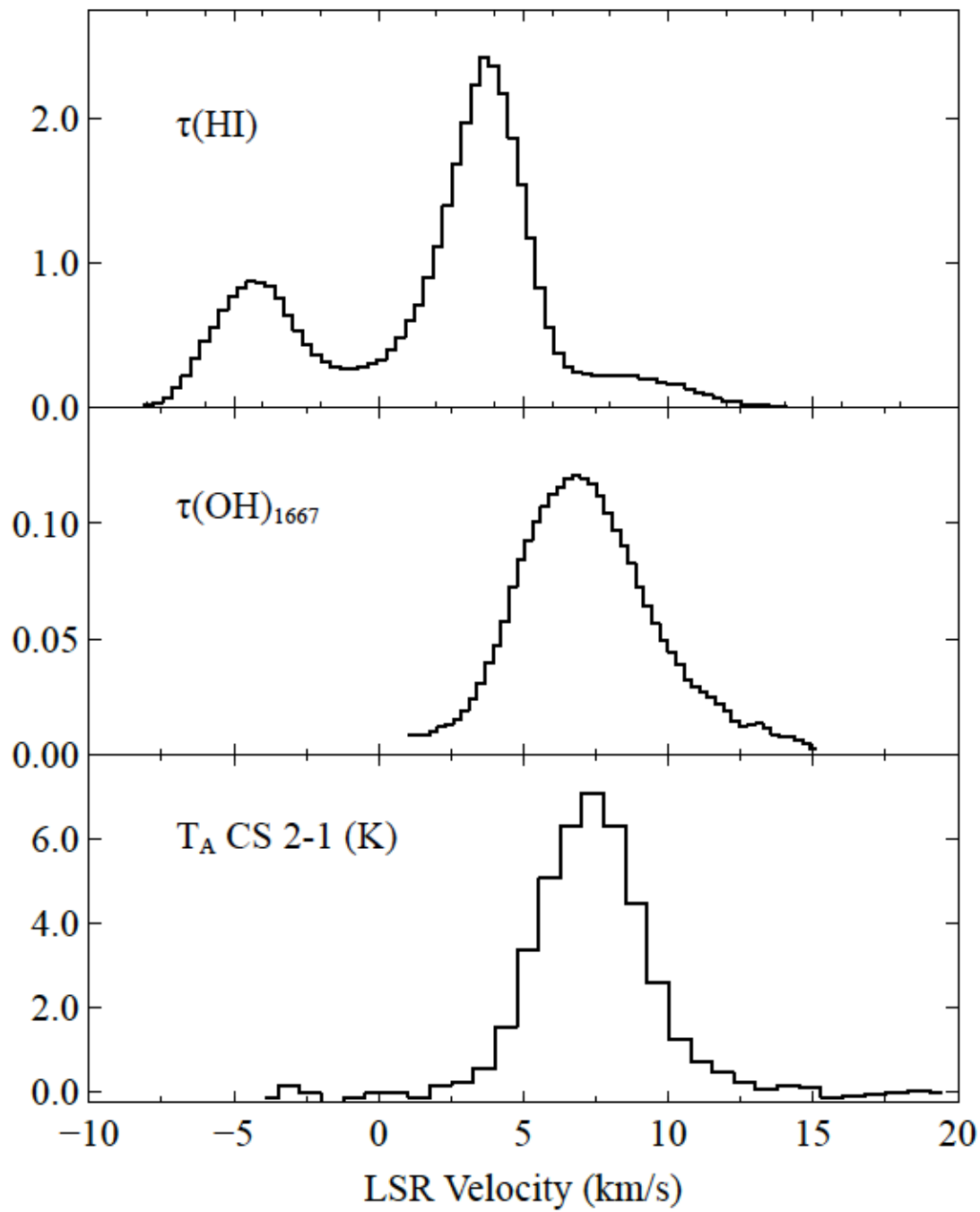

**Figure 10**. Profiles toward center of Orion-S for H I optical depth (top), 1667 MHz OH optical depth (middle), and CS (2-1) emission (bottom). H I and OH profiles come from the present work. See Tables 1a and 1b for beamwidths and channel separations. CS profile is from Tatematsu et al. (1998); beamwidth is 50", channel separation is 0.75 km s$^{-1}$.

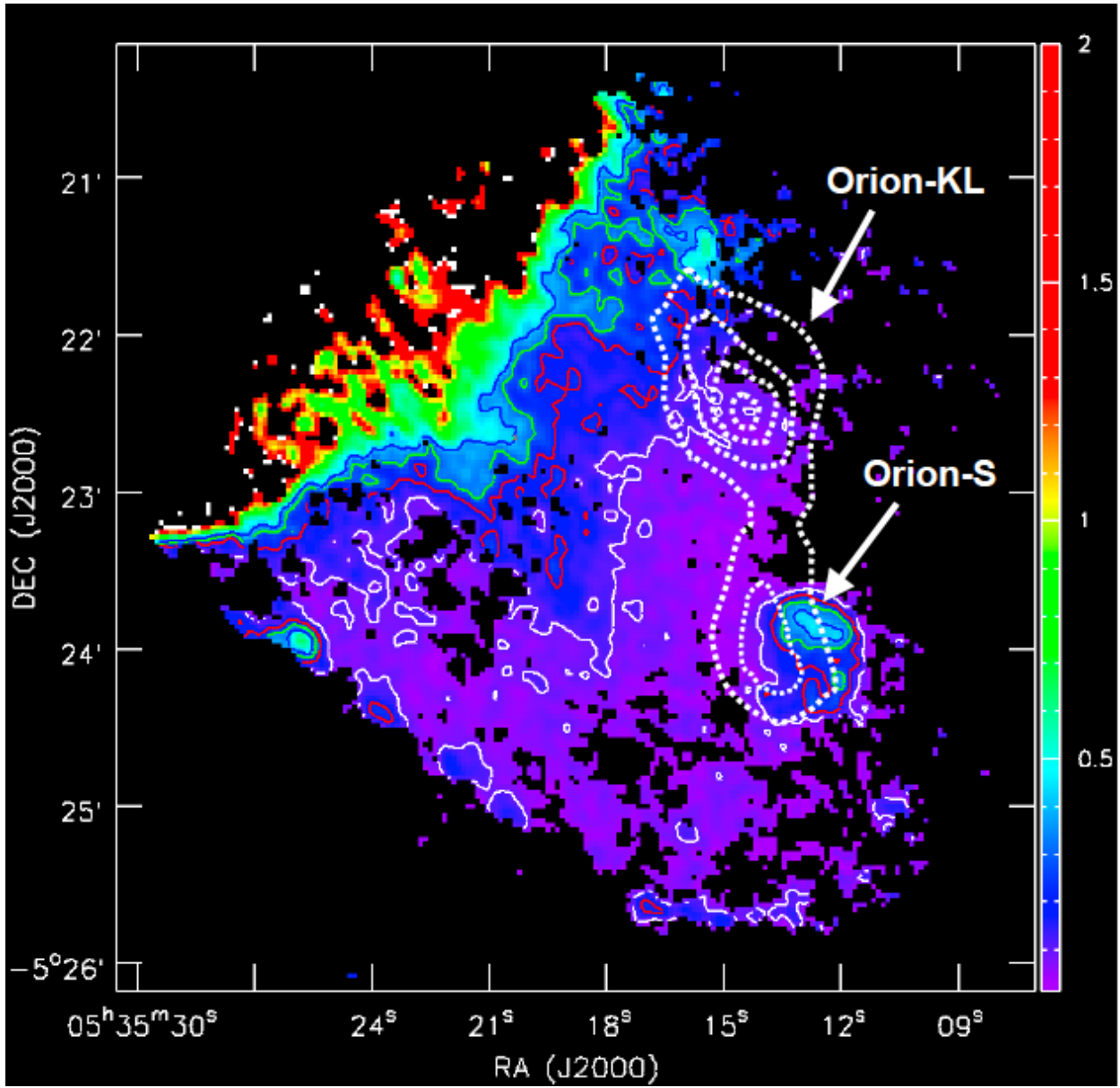

**Figure 11.** H I optical depth averaged over the velocity range 8.4 to 9.7 km s$^{-1}$, taken from data of vdWGO13 (colors and light contours). Synthesized beamwidth is 6.4". Optical depth contours are for 0.1, 0.2, 0.3 and 0.4 napiers. Pixels with indeterminate optical depth have been set to zero. Overlaid are contours for 850 μm JCMT/SCUBA dust continuum emission from data of Johnstone & Bally (1999), as plotted by Peng et al. (2012). Dust emission contours (heavy, dashed lines) are for 7%, 20%, 40%, and 80% of the peak flux of 87 Jy beam$^{-1}$, where the beamwidth is 14". Locations of Orion-KL and Orion-S in the 850 μm image are indicated.

# TABLE 1a
Observational parameters of 21cm VLA HI observations of Orion-A (M42) Region

| Parameter | Value |
|---|---|
| Frequency | 1420.4 MHz |
| Observing dates (project code, configuration) | 1991 Jan 20, Jan 21 (AT113, C) |
| Total observing time on source | 10.1 hr |
| Primary beam HPBW | 30' |
| Synthesized beam HPBW | 25" × 25" |
| Phase and pointing center (J2000) | $05^h35^m17^s.5$, $-05°23'37"$ |
| Primary flux density calibrator | 3C48 |
| Phase calibrator | 0607-085 (J2000) |
| Frequency channels per polarization | 128 (Two IF mode) |
| Total bandwidth | 195.3 kHz (41.2 km s$^{-1}$) |
| Velocity coverage (LSR) | -17 to +24 km s$^{-1}$ |
| Channel separation | 1.53 kHz (0.32 km s$^{-1}$) |
| Channel resolution[a] | 1.83 kHz (0.39 km s$^{-1}$) |
| Pixel size in images | 4" × 4" |
| Rms noise in line channels | 8.5 mJy beam$^{-1}$ |
| Rms noise in continuum | 16 mJy beam$^{-1}$ |
| $S_v$ to $T_b$ | 1 mJy beam$^{-1}$ = 1.1 K |
| Angular to linear scale[b] | 1' = 0.13 pc |

[a] After Hanning smoothing
[b] Assumes distance of 450 pc.

# TABLE 1b

Observational parameters of 18 cm VLA OH observations of Orion-A (M42) Region

| Parameter | Value |
|---|---|
| Frequencies | 1665.4 and 1667.4 MHz |
| Observing date (project code, configuration) | 1997 Nov 20 (AT209, D) |
| Total observing time on source | 5.9 hr |
| Primary beam HPBW | 26' |
| Synthesized beam HPBW | 47" × 39", PA = -23° |
| Phase and pointing center (J2000) | $05^h35^m17^s.5$, $-05°23'37"$ |
| Primary flux density calibrators | 3C48, 3C147 |
| Phase calibrator | 0503+020 (J2000) |
| Frequency channels per polarization | 128 (Four IF Mode) |
| Total bandwidth | 195.3 kHz (35.2 km s$^{-1}$) |
| Velocity coverage (LSR) | -12 to +24 km s$^{-1}$ |
| Channel separation | 1.53 kHz (0.27 km s$^{-1}$) |
| Channel resolution | 1.83 kHz (0.33 km s$^{-1}$) |
| Pixel size in images | 10" × 10" |
| Rms noise in line channels | 4.3 mJy beam$^{-1}$ |
| Rms noise in continuum | 9 mJy beam$^{-1}$ |
| $S_v$ to $T_b$ | 1 mJy beam$^{-1}$ = 0.27 K |
| Angular to linear scale[a] | 1' = 0.13 pc |

[a]Assumes distance of 450 pc.

# TABLE 2
Parameters for positions with 18 cm OH absorption toward Orion-A

| No. | Position location | R.A.[a] (J2000) | Decl.[a] (J2000) | $V_{o,67}$[b] (km s$^{-1}$) | $\Delta V_{67}$[c] (km s$^{-1}$) | $\tau_{o,67}$[d] | $\tau_{o,67}/\tau_{o,65}$[e] | N(OH)[f] (cm$^{-2}$) | $A_v$[g] (mag) | N(H)[h] (cm$^{-2}$) |
|---|---|---|---|---|---|---|---|---|---|---|
| OH1 | Northeast Dark Lane 1 | 05$^h$35$^m$32$^s$.2 | -05°22'06" | 6.4 | 2.7 | 0.37 | 1.7 | 4.7×10$^{15}$ | 58 | 1.5×10$^{23}$ |
| OH2 | Northeast Dark Lane 2 | 05$^h$35$^m$31$^s$.4 | -05°20'46" | 6.6 | 1.9 | 0.97 | 2.2 | 8.8×10$^{15}$ | 110 | 2.7×10$^{23}$ |
| OH3 | Dark Bay | 05$^h$35$^m$23$^s$.4 | -05°22'29" | 3.7 | 2.7 | 0.04 | 1.9 | 5.6×10$^{14}$ | 7 | 1.7×10$^{22}$ |
| OH4 | Orion-S | 05$^h$35$^m$12$^s$.1 | -05°23'56" | 6.9 | 5.2 | 0.12 | 1.2 | 3.8×10$^{15}$ | 48 | 1.2×10$^{23}$ |
| OH5 | M43 | 05$^h$35$^m$37$^s$.7 | -05°15'46" | 9.6 | 2.2 | 0.59 | 1.9 | 5.8×10$^{15}$ | 72 | 1.8×10$^{23}$ |

[a]Position of the center of the 40" synthesized beam at 18 cm

[b]Center velocity of 1667 MHz OH line

[c]FWHM width of 1667 MHz OH line

[d]Peak optical depth of 1667 MHz OH line

[e]Ratio of peak optical depths of 1667 and 1665 MHz OH lines. LTE value is 1.8.

[f]OH column density, taken as the average of OH column densities computed separately from 1665 and 1667 MHz OH lines. Calculation of OH column densities from the individual OH lines was done with the relationships of Roberts et al. (1995), assuming $T_{ex}$ = 20 K.

[g]$A_v$ estimated from the ratio N(OH)/Av = 8 × 10$^{13}$ cm$^{-2}$ mag$^{-1}$ of Crutcher (1979) who states that the ratio is valid over the range $A_v$ = 0.4 to 7 mag. If the ratio N(OH)/Av decreases for $A_v$ > 7 mag, owing to freeze-out or chemical conversion of OH to other species, then some values of $A_v$ in this column are underestimates.

[h]N(H) estimated from the ratio $A_v$/N(H) = 4 × 10$^{-22}$ mag cm$^2$ cited by Abel et al. (2004) for the Orion region. Some of these values could be underestimates, see note regarding $A_v$ immediately above.

# TABLE 3
Parameters for selected positions with unsaturated HI absorption toward Orion-A

| No. | R.A.[a] (J2000) | Decl.[a] (J2000) | HI Component A | | | | HI Component B | | | |
|---|---|---|---|---|---|---|---|---|---|---|
| | | | $V_o$[b] (km s$^{-1}$) | $\Delta V$[c] (km s$^{-1}$) | $N(H)$[d] (cm$^{-2}$) | $B_{los}$ ($\mu$G) | $V_o$[b] (km s$^{-1}$) | $\Delta V$[c] (km s$^{-1}$) | $N(H)$[e] (cm$^{-2}$) | $B_{los}$ ($\mu$G) |
| HI1 | 05$^h$ 35$^m$ 16$^s$.7 | -05° 23' 25" | 5.3 | 2.3 | 1.9×10$^{21}$ | -52±4 | 1.0 | 3.2 | 2.7×10$^{21}$ | -47±4 |
| HI2 | 05$^h$ 35$^m$ 16$^s$.7 | -05° 22' 37" | 5.8 | 2.3 | 2.0×10$^{21}$ | -50±5 | 2.3 | 3.6 | 3.1×10$^{21}$ | -35±7 |
| HI3 | 05$^h$ 35$^m$ 16$^s$.6 | -05° 24' 13" | 5.0 | 2.6 | 1.7×10$^{21}$ | -40±4 | -0.4 | 3.2 | 1.7×10$^{21}$ | -64±6 |
| HI4 | 05$^h$ 35$^m$ 19$^s$.9 | -05° 23' 25" | 5.5 | 3.0 | 2.7×10$^{21}$ | -60±5 | 2.5 | 3.9 | 4.1×10$^{21}$ | -43±6 |
| HI5 | 05$^h$ 35$^m$ 13$^s$.5 | -05° 23' 24" | 4.4 | 2.4 | 1.3×10$^{21}$ | -48±7 | n/a[f] | n/a[f] | n/a[f] | n/a[f] |
| HI6[h] | 05$^h$ 35$^m$ 12$^s$.1 | -05° 23' 56" | 3.8 | 3.2 | 1.3×10$^{21}$ | -27±15 | n/a[f] | n/a[f] | n/a[f] | n/a[f] |
| HI7 | 05$^h$ 35$^m$ 14$^s$.2 | -05° 24' 45" | 5.0 | 2.9 | 2.0×10$^{21}$ | -79+7 | 0.6 | 3.9 | 1.2×10$^{21}$ | -35±10 |
| HI8 | 05$^h$ 35$^m$ 20$^s$.9 | -05° 24' 25" | 5.2 | 1.9 | 1.2×10$^{21}$ | -31±5 | n/a[f] | n/a[f] | n/a[f] | n/a[f] |
| HI9 | 05$^h$ 35$^m$ 21$^s$.5 | -05° 23' 45" | 5.7 | 2.4 | 1.7×10$^{21}$ | -73±7 | 1.8 | 3.6 | 2.6×10$^{21}$ | n/a[g] |

[a] Position of the center of the 25" synthesized beam at 21 cm

[b] Center velocity of HI line component

[c] FWHM width of HI line component

[d] $N(H)$ computed assuming $T_{ex}$ = 90 K (Abel et al. 2006)

[e] $N(H)$ computed assuming $T_{ex}$ = 135 K (Abel et al. 2006)

[f] HI component not clearly identifiable in the profile.

[g] Fitted value of $B_{los}$ does not meet criterion of $|B_{los}|/\sigma(B_{los}) > 3$.

[h] Position coincides with Orion-S; however, velocity, $N(H)$ and $B_{los}$ values are for the unrelated HI Component A.

# TABLE 4a
Estimated energy parameters for HI gas sampled by 21 cm HI line

| Parameter[a] | HI Component A (SW Sector)[b] | HI Component B (SW Sector)[c] | CNM[d] |
|---|---|---|---|
| $M_{turb}$[e] | 2.7 | 3.8 | 3.0 |
| $\beta_{therm}$[f] | 0.007 | 0.10 | 0.29 |
| $\beta_{turb}$[g] | 0.017 | 0.46 | 0.88 |
| $\lambda$[h] | 0.09 | 0.14 | 0.08 |
| $E_{therm}$ (%)[i] | 1 | 8 | 16 |
| $E_{turb}$ (%)[i] | 3 | 38 | 48 |
| $E_B$ (%)[i] | 96 | 54 | 36 |
| $E_{therm}+E_{turb}+E_B$[j] | $3.2 \times 10^{-10}$ | $4.4 \times 10^{-10}$ | $3.9 \times 10^{-12}$ |

[a]In computing energy parameters involving magnetic fields, we take $B_{tot} = 2B_{los}$ and $B_{tot}^2 = 3B_{los}^2$ (see Crutcher 1999).

[b]Energy parameters are computed from averages over positions 1-9 of relevant parameters ($\Delta V$, $B_{los}$, $N(H)$) in Table 3. Also, we take $n(H) = 300$ cm$^{-3}$ and $T_K = 50$ K (Abel et al. 2015). All positions included in the averages lie in the southwest sector of the Veil where the HI line is unsaturated.

[c]Energy parameters are computed from averages over positions 1-4 and 7 of relevant parameters ($\Delta V$, $B_{los}$, $N(H)$) in Table 3. Also, we take $n(H) = 2500$ cm$^{-3}$ and $T_K = 60$ K (Abel et al. 2015). All positions included in the averages lie in the southwest sector of the Veil where the HI line is unsaturated.

[d]Energy parameters for the Cold Neutral Medium (CNM) are computed from mean values of relevant parameters ($\Delta V$, $n(H)$, $N(H)$, $T_K$ and $B_{tot} = 6$ μG) given by Heiles & Troland (2005).

[e]*Turbulent Mach number*, measure of turbulent to thermal energy densities, $(1/3) M_{turb}^2 = E_{turb}/E_{therm}$.

[f]*Conventional plasma parameter*, measure of thermal to magnetic energy densities, $\beta_{therm} = (2/3) E_{therm}/E_B$.

[g]*Turbulent plasma parameter*, measure of turbulent to magnetic energy densities, $\beta_{turb} = (2/3) E_{turb}/E_B$.

[h]*Dimensionless mass-to-flux ratio*, a measure of the gravitational to magnetic energies.

[i]*Per cent* of ($E_{therm} + E_{turb} + E_B$) represented by the specified energy.

[j]Ergs cm$^{-3}$

# TABLE 4b
Estimated energy parameters for molecular gas sampled via 18 cm OH lines

| Parameter[a] | Orion Veil | | | | | Low mass molecular cores[b] | High mass molecular cores | | |
|---|---|---|---|---|---|---|---|---|---|
| | Northeast Dark Lane 1 | Northeast Dark Lane 2 | Dark Bay | Orion-S | M43 | Mean values | M17 core B17S[c] | S106[d] | S88B[e] |
| $M_{turb}$ | 6 | 4 | 6 | 12 | 5 | 2 | 5 | 5 | 6 |
| $\beta_{therm}$ | 0.07 | n/a[f] | n/a[f] | n/a[f] | n/a[f] | 1.3 | 0.12 | 0.03 | 0.16 |
| $\beta_{turb}$ | 0.81 | n/a[f] | n/a[f] | n/a[f] | n/a[f] | 1.5 | 0.96 | 0.17 | 1.7 |
| $\lambda$ | 1.1 | n/a[f] | n/a[f] | n/a[f] | n/a[f] | 2.8 | 8.5 | 1.4 | 4.7 |
| $E_{therm}$ (%)[g] | 4 | n/a[f] | n/a[f] | n/a[f] | n/a[f] | 38 | 7 | 3 | 6 |
| $E_{turb}$ (%)[g] | 52 | n/a[f] | n/a[f] | n/a[f] | n/a[f] | 43 | 55 | 20 | 67 |
| $E_B$ (%)[g] | 43 | n/a[f] | n/a[f] | n/a[f] | n/a[f] | 19 | 38 | 77 | 27 |
| $E_{therm}+E_{turb}+E_B$[h] | $3.4 \times 10^{-8}$ | n/a[f] | n/a[f] | n/a[f] | n/a[f] | $4.2 \times 10^{-11}$ | $2.8 \times 10^{-8}$ | $2.5 \times 10^{-8}$ | $1.8 \times 10^{-8}$ |

[a] See notes a, c, f, g, h, i and j to Table 4a for parameter specifications. At Northeast Dark Lane 1 position, $B_{los}$ = -350 µG; see section 3.5. $T_K$ assumed to be 30 K for OH associated with massive star formation (Orion Veil and high mass molecular cores), based on Orion photoionization model of O'Dell et al. (2009). $T_K$ assumed to be 20 K for low mass molecular cores. Volume density for Northeast Dark Lane 1 position (necessary for computing $\beta_{therm}$ and $\beta_{turb}$) estimated as $4 \times 10^5$ cm$^{-3}$, based on $N(H)$ from Table 2 and the dimension in plane of the sky (0.12 pc) of OH feature.

[b] Mean values of $\Delta V$, $B_{los}$, $N(H)$ and $n(H)$ taken for 34 molecular cores observed by Troland & Crutcher (2008) in OH emission. All but four of these cores are associated with low-mass star formation and have masses in range 5 to 100 $M_{solar}$.

[c] Values of $\Delta V$, $B_{los}$, $N(H)$ and $n(H)$ based upon OH absorption by molecular core B17S of Brogan & Troland (2001). We assume $T_{ex}$ = 20 K for the OH lines. The B17S core has $M \approx 4000\ M_{solar}$.

[d] Same as above for M17 except for the core associated with S106 OH velocity component B observed by Roberts, Crutcher & Troland (1995). This core has $M \approx 250\ M_{solar}$.

[e] Same as above for M17 except for core in S88B observed by Sarma et al. (2013). This core has $M \approx 600\ M_{solar}$.

[f] Value of parameter unknown since the magnetic field was not detected.

[g] *Per cent* of ($E_{therm} + E_{turb} + E_B$) represented by the specific energy; these parameters only computed for positions where all three energies can be estimated.

[h] Ergs cm$^{-3}$